\documentclass[fleqn,usenatbib]{mnras}

\usepackage{newtxtext,newtxmath}

\usepackage[T1]{fontenc}
\usepackage[pagewise]{lineno}

\DeclareRobustCommand{\VAN}[3]{#2}
\let\VANthebibliography\thebibliography
\def\thebibliography{\DeclareRobustCommand{\VAN}[3]{##3}\VANthebibliography}

\newcommand{\kms}{\ensuremath{\mathrm{km}\,\mathrm{s}^{-1}}}

\newcommand{\Msun}{\ensuremath{\rm{{M}_{\odot}}}}

\usepackage{graphicx}	
\usepackage{amsmath}	
\hypersetup{
  colorlinks,
  citecolor=blue,
  linkcolor=blue,
  urlcolor=magenta}
\usepackage{orcidlink}

\defcitealias{2024ApJ...961..161M}{M24}
\defcitealias{marchal_2019}{M19}

\title[H{\sc i} shells and loops in the Magellanic Bridge]{A Detailed GASKAP-HI View of Shells and Loops in the Magellanic Bridge}

\author[S.-J. Kim et al.]{Shin-Jeong Kim\,\orcidlink{0000-0002-6760-7531},$^{1}$\thanks{E-mail: Shin-Jeong.Kim@anu.edu.au} 
Antoine Marchal\,\orcidlink{0000-0002-5501-232X},$^{2,1}$ 
N.\ M.\ McClure-Griffiths\,\orcidlink{0000-0003-2730-957X},$^{1,3}$
J. R. Dawson\,\orcidlink{0000-0003-0235-3347},$^{4,5}$
\newauthor
James Dempsey\,\orcidlink{0000-0002-4899-4169},$^{1}$
Helga Dénes\,\orcidlink{0000-0002-9214-8613},$^{7}$
John M. Dickey\,\orcidlink{0000-0002-6300-7459},$^{8}$
Steven J. Gibson\,\orcidlink{0000-0002-1495-760X},$^{9}$  
Katie Jameson\,\orcidlink{0000-0001-7105-0994},$^{10}$  
\newauthor
Ian Kemp\,\orcidlink{0000-0002-6637-9987},$^{11,12}$  
Bumhyun Lee\,\orcidlink{0000-0002-3810-1806},$^{13}$
Min-Young Lee\,\orcidlink{0000-0002-9888-0784},$^{14}$
Adam K. Leroy\,\orcidlink{0000-0002-2545-1700},$^{15}$ 
Callum Lynn\,\orcidlink{0000-0001-6846-5347},$^{1}$  
\newauthor
Yik Ki Ma\,\orcidlink{0000-0003-0742-2006},$^{1,16}$  
Marc-Antoine Miville-Deschênes\,\orcidlink{0000-0002-7351-6062},$^{2}$  
Eric G. M. Muller\,\orcidlink{0000-0001-5621-1577},$^{1}$
Claire Murray\,\orcidlink{0000-0002-7743-8129},$^{17}$ 
\newauthor
Hiep Nguyen\,\orcidlink{0000-0002-2712-4156},$^{1}$ 
Nickolas Pingel\,\orcidlink{0000-0001-9504-7386},$^{18,19}$
Hye-Jin Park\,\orcidlink{0000-0002-9809-6631},$^{1}$  
and Jacco Th. van Loon\,\orcidlink{0000-0002-1272-3017}$^{20}$   \\
 \\
$^{1}$Research School of Astronomy $\&$ Astrophysics, Australian National University, Canberra ACT 2611, Australia\\
$^{2}$Laboratoire de Physique de l’École Normale Supérieure, ENS, Université PSL, CNRS, Sorbonne Université, Université Paris Cité, Observatoire de Paris, \\ F75005 Paris, France\\
$^{3}$SKA Observatory, Jodrell Bank, Lower Withington, Macclesfield, SK11 9FT, UK\\
$^{4}$School of Mathematical and Physical Sciences and Astrophysics and Space Technologies Research Centre, Macquarie University, NSW 2109, Australia\\
$^{5}$CSIRO Space $\&$ Astronomy, Australia Telescope National Facility, P.O. Box 76, Epping, NSW 1710, Australia\\
$^{7}$College of Sciences and Engineering, Universidad San Francisco de Quito, Quito, 170901, Ecuador \\
$^{8}$School of Natural Sciences, University of Tasmania, Hobart, TAS 7005, Australia\\
$^{9}$Department of Physics and Astronomy, Western Kentucky University, Bowling Green, KY 42101, USA\\
$^{10}$Caltech Owens Valley Radio Observatory, Pasadena, CA 91125, USA\\
$^{11}$International Centre for Radio Astronomy Research (ICRAR), Curtin University, Bentley, WA 6102, Australia\\
$^{12}$CSIRO Space and Astronomy, 26 Dick Perry Avenue, Kensington, 6151, WA, Australia\\
$^{13}$Department of Astronomy, Yonsei University, 50 Yonsei-ro, Seodaemun-gu, Seoul 03722, Republic of Korea\\
$^{14}$Korea Astronomy and Space Science Institute, 776 Daedeokdae-ro, Yuseong-gu, Daejeon 34055, Republic of Korea\\
$^{15}$Department of Astronomy, The Ohio State University, 140 West 18th Avenue, Columbus, OH 43210, USA\\
$^{16}$Max-Planck-Institut f\"ur Radioastronomie, Auf dem H\"ugel 69, 53121 Bonn, Germany \\ 
$^{17}$Space Telescope Science Institute, 3700 San Martin Drive, Baltimore, MD 21218, USA  \\
$^{18}$Department of Astronomy, University of Wisconsin--Madison, 475 N Charter St., Madison, WI 53706, USA \\
$^{19}$Department of Astronomy, Indiana University, 727 East Third Street, Bloomington, IN 47405, USA\\
$^{20}$Lennard-Jones Laboratories, Keele University, ST5 5BG, UK} 

\date{Accepted XXX. Received YYY; in original form ZZZ}

\pubyear{\the\year{}}

\usepackage[font=small,skip=2pt]{caption}
\usepackage{titlesec}
\titlespacing*{\section}{0pt}{8pt}{4pt}
\titlespacing*{\subsection}{0pt}{6pt}{3pt}

\begin{document}
\label{firstpage}
\pagerange{\pageref{firstpage}--\pageref{lastpage}}
\maketitle

\begin{abstract}
We present 8\,pc-scale GASKAP-H{\sc i} observations of an H{\sc i} shell (diameter $\sim$130\,pc) associated with the H$\alpha$ shell DEM171 in the Magellanic Bridge.
The Bridge’s diffuse, H{\sc i}-dominated environment, with minimal galactic shearing and fewer overlapping star-forming regions, provides an ideal environment to study cold gas formation driven by stellar feedback.
To investigate the H{\sc i} multi-phase structure and kinematics of the shell, we perform Gaussian decomposition of H{\sc i} line profiles.
We identify cold H{\sc i} components with velocity dispersions $\sigma < 2.5\,\kms$ located within the shell. 
Complementary Herschel far-infrared (FIR) 250$\micron$ and H$\alpha$ maps show that while the central cavity is ionized, the shell walls contain cold H{\sc i}, molecular gas, and dust, suggesting that stellar feedback has promoted cold gas formation or swept pre-existing cold gas into the shell walls.
We also identify cold H{\sc i} clumps outside the shell tracing a larger-scale structure, MB-Loop I.
To investigate the large-scale context, we apply a Fourier transform method to H{\sc i} emission-line profiles to map the lower limit of cold-gas column densities across a section of the Bridge, revealing structures from large loops to smaller shells.
The association between MB-Loop~I and the H{\sc i} shell suggests that cold H{\sc i} gas may pre-exist before shell expansion.
Finally, the shell exhibits asymmetric expansion, with a preferred orientation roughly perpendicular to the arc along MB-Loop I.
Our results show that cold H{\sc i} gas exists across a wide range of spatial scales in the Magellanic Bridge, highlighting the dynamic interplay that shapes the surrounding interstellar medium.

\end{abstract}




\section{Introduction}

\label{sec:intro} 
The interstellar medium (ISM) consists of gas and dust. The bulk of the gas mass in late-type galaxies is in the form of neutral hydrogen (H{\sc i}, \citealt{2022ARA&A..60..319S}), while the star-forming gas is molecular hydrogen (H$_2$, \citealt{2011AJ....142...37S}). The H{\sc i} gas consists of two primary stable phases: the diffuse warm neutral medium (WNM) and the dense cold neutral medium (CNM), as well as the unstable neutral medium (UNM), which is in the process of transitioning between these two phases (\citealt{2023ARA&A..61...19M}). The spatial distribution and phase balance of the H{\sc i} gas are set by a complex interplay between the dynamical processes at play \citep[e.g., turbulence, stellar feedback, colliding flows;][]{Elmegreen:2004, Scalo:2004} and the balance between the cooling and heating processes acting in the ISM (\citealt{1995ApJ...443..152W, 2003ApJ...587..278W}).

In particular, stellar feedback, such as supernovae (SNe) or stellar winds from massive stars, plays a crucial role in shaping the ambient ISM (\citealt{2024ARA&A..62..369S}). It injects a large amount of energy and momentum, creating expanding structures known as bubbles or shells ranging from a few tens of parsecs to kiloparsec scales. While these dynamic events can destroy pre-existing molecular clouds (e.g., \citealt{2011ApJ...741...85D, 2022MNRAS.509..272C}), they can also compress ambient gas within the shells, leading to the formation of new molecular clouds and potentially triggering subsequent star formation (\citealt{1977ApJ...214..725E, 1987ApJ...317..190M, 1994MNRAS.268..291W, 2011ApJ...728..127D, 2013ApJ...763...56D, 2013PASA...30...25D, 2014ApJ...796..123F, 2015ApJ...799...64D, 2017MNRAS.472.2975M}). 

Simulations have shown that the formation of the CNM from diffuse warm gas occurs through the thermal condensation driven by converging gas flows and/or turbulence (\citealt{2005A&A...433....1A, 2008ApJ...687..303I, 2014A&A...567A..16S, 2020ApJ...905...95K}). Similarly, \citet{2011ApJ...731...13N} models showed that cold (T < 100\,K) and dense clumps form at the compressed region between colliding wind-blown superbubbles. Because CNM is an important precursor to dense molecular clouds and star formation, exploring the mechanisms of its formation within feedback-driven structures like shells is crucial for understanding the broader process of star formation and galaxy evolution.

Observational studies have catalogued numerous H{\sc i} shells in the Milky Way and nearby galaxies  (\citealt{1997MNRAS.289..225S, 1999AJ....118..273W, 1999AJ....118.2797K, 2002ApJ...578..176M, 2011AJ....141...23B, 2014A&A...564A.116S, 2020AJ....160...66P}). Notably, high-resolution observations have further revealed the presence of cold H{\sc i} gas at the shell walls. \citet{2003ApJ...594..833M} found small-scale structures with narrow H{\sc i} linewidths ($1.5-2.5$\,\kms) in the Galactic supershell GSH 277+00+36. \citet{2011ApJ...728..127D} investigated two Galactic supershells, GSH 287+04-17 and GSH 277+00+36, and revealed that narrow ($\sim$3\,\kms) and cold ($\sim$100\,K) gas structures are present in both shell walls. The H{\sc i} self-absorption (HISA) feature has also been detected in the Galactic supershell GSH 006−15+7, indicating the presence of cold atomic gas, as HISA occurs when cold atomic gas lies in front of the warmer background gas (\citealt{2012MNRAS.421.3159M}). Additionally, direct H{\sc i} absorption observations reveal CNM gas toward a supergiant shell in the Large Magellanic Cloud (LMC) (\citealt{2000A&A...354..787M}). 
Despite these observations, our understanding of the multi-phase structure of H{\sc i} shells, particularly the distribution of CNM and WNM, remains limited.
 
To better understand the multi-phase H{\sc i} properties within shell structures, we investigate the Magellanic Bridge.
The Magellanic Bridge is a tidally stripped gas structure connecting the Small Magellanic Cloud (SMC) and LMC. It is suggested to have formed $\sim$200\,Myr ago by the close tidal interaction/encounter between the SMC and LMC (\citealt{1994MNRAS.266..567G, 2012MNRAS.421.2109B, 2022MNRAS.515..940W}). The Magellanic Bridge offers a close ($\sim$56\,kpc\footnote{Throughout this paper, we assume a distance of 56\,kpc to the Magellanic Bridge, taken as the mean of the distances to the LMC ($\sim$50\,kpc; \citealt{2014AJ....147..122D}) and the SMC ($\sim$62\,kpc; \citealt{2015AJ....149..179D}).}) environment, where the effects of the galactic shearing or differential rotation are not significant and the volume density is low, allowing shell structures to grow larger and persist for longer time-scales (\citealt{1987ApJ...317..190M, 1999AJ....118..273W}). 
%
Nevertheless, in these conditions, star formation occurs in some locally denser regions along the Bridge. Spatially clustered young stars along the densest H{\sc i} gas regions in the Bridge imply in-situ star formation (\citealt{2017MNRAS.472.2975M}). Such findings are further supported by observations of young stellar objects (YSOs, \citealt{2014ApJ...785..162C, 2020MNRAS.499.2534K}) and O- and B-type stars in the Bridge (\citealt{1999A&A...348..728R, 2021A&A...646A..16R}), marking active and in-situ star formation (e.g., \citealt{2014ApJ...785..162C, 2021A&A...646A..16R}). Additionally, the detection of ionized gas traced by H$\alpha$ emission (\citealt{1986MNRAS.223..317M, 2005MNRAS.362..689P, 2007PASA...24...69M}), and molecular gas traced by CO line emission (\citealt{2003MNRAS.338..609M, 2006ApJ...643L.107M, 2014PASJ...66....4M, 2020MNRAS.499.2534K, 2020A&A...641A..97V}) demonstrates the potential for ongoing star formation in the Bridge. Given this diffuse environment (\citealt{2003MNRAS.339..105M, 2008ApJ...678..219L}), the formation of CNM might represent a critical bottleneck regulating star formation in this region.

\citet{2003MNRAS.339..105M} revealed that the Magellanic Bridge consists of filaments, arcs, loops, and shell structures in H{\sc i}, indicating its dynamic environment. They also identified 163 candidate shells, with mean radii of 60\,pc, by visually examining three projections (RA--Dec., RA--velocity, and Dec.--velocity) of the H{\sc i} data cube. 
Among their catalogue, we focus on H{\sc i} Shell-91, which shows a clear spatial correspondence with the giant H$\alpha$ shell DEM171 and therefore provides a well-defined feedback-driven structure.

The H$\alpha$ shell DEM171 was first identified by \citet{1986MNRAS.223..317M} in an H$\alpha$ survey targeting the SMC `wing'. 
Reporting a radius of $4'$ ($\sim$\,80\,pc at the distance of the SMC), they suggested a stellar wind from an O star or a supernova remnant (SNR) origin, with a possible age of 5\,Myr.
%
%
Later, \citet{Parker1998} identified an ultraviolet (UV) source--FAUST 392, located near this shell, and listed the possible ionizing source candidates as a planetary nebula, SNR, or Wolf-Rayet (WR) star.
In a later H$\alpha$ study, \citet{2001MNRAS.326..539G} measured an expansion velocity of 37\,\kms for DEM171 and proposed a WR star as its possible ionizing source. 
While previous optical surveys placed this shell as part of the SMC `wing', the H{\sc i} study by \citet{2003MNRAS.339..105M} placed it within the Magellanic Bridge, and identified the H{\sc i} shell-91 as the counterpart of the H$\alpha$ DEM171 shell.
\citet{2003MNRAS.339..105M} measured a radius of $4.5'$ ($\approx$80\,pc at their assumed distance of 60\,kpc), a dynamical age of $9\pm2$\,Myr, and an H{\sc i} expansion velocity of $5\pm2$\,\kms.
This expansion velocity is comparable to the sound speed of the warm H{\sc i} (\citealt{2023ARA&A..61...19M}). 
This H{\sc i} shell, also traced in H$\alpha$ emission, provides an ideal laboratory to study the effects of stellar feedback on H{\sc i} gas properties, including gas kinematics and the formation of the CNM. 
In particular, we investigate how cold H{\sc i} gas is structured across multiple spatial scales, from the internal structure of Shell-91 to its immediate multi-phase environment, and up to the larger-scale cold gas distribution in the Magellanic Bridge.

In this paper, we present high-resolution (30\arcsec\,, corresponding to an 8\,pc scale) images of the H{\sc i} shell in the Magellanic Bridge observed with the Galactic Australian Square Kilometre Array Pathfinder (GASKAP; \citealt{2013PASA...30....3D}; \citealt{2022PASA...39....5P}) survey. 
This paper is organized as follows. In Section~\ref{sec:data}, we describe the data used in our analysis.
In Section~\ref{sec:ROHSA}, we present a Gaussian decomposition of the H{\sc i} emission data in Shell-91 using {\tt ROHSA} \citep{marchal_2019}, and analyse the distribution of cold gas components as well as the overall dynamics of the shell. 
In Section~\ref{sec:FFT}, we apply a Fourier Transform (hereafter FT) method \citep{2024ApJ...961..161M} to map a lower limit of the cold H{\sc i} column density across a section of the Magellanic Bridge to reveal the large-scale context.
A discussion and a summary are presented in Sections~\ref{sec:discussion} and \ref{summary}, respectively.

\section{GASKAP-H{\sc i} observations}
\label{sec:data} 
In this work, we use H{\sc i} 21\,cm emission data of the Magellanic Bridge from the GASKAP-H{\sc i} pilot observations, specifically from scheduling block (SB) 14180, observed on 2020 May 13 during Pilot Phase I \citep[for details, see][]{2026MNRAS.549ag863D}. 
GASKAP-H{\sc i} aims to explore the multi-phase ISM and its role in star formation through observations of neutral hydrogen at 21\,cm in the Milky Way and Magellanic System. 
The observations were made with the ASKAP telescope, consisting of 36$\times$12\,m dishes located in Western Australia (\citealt{2021PASA...38....9H}).
The resulting H{\sc i} data cube has dimensions of 2892$\times$2880 pixels, with a pixel scale of 7\arcsec\,and 181 velocity channels with 0.98\,\kms\, velocity resolution.
The pilot observations toward the Magellanic Bridge were conducted with a total integration time of 10\,h, yielding an rms noise level of 1.1\,K per 0.98\,\kms velocity channel (\citealt{2022PASA...39....5P}).
The imaging process, including the combination with Galactic All-Sky Survey (GASS; \citealt{2009ApJS..181..398M, 2010A&A...521A..17K}) Parkes single-dish data, is discussed in \citet[see their Sections 3.3.2 and 3.3.3]{2022PASA...39....5P}. The data used here have an angular resolution of 30\arcsec\,and a spectral resolution of 0.98\,\kms, resolving gas structures at about 8\,pc scale at the distance of the Magellanic Bridge ($\sim$56\,kpc).

\section{H{\sc i} view of Shell-91}\label{sec:ROHSA}
\subsection{PPV-space structure}\label{sec:3.1}
\begin{figure}
    \includegraphics[width=\columnwidth]{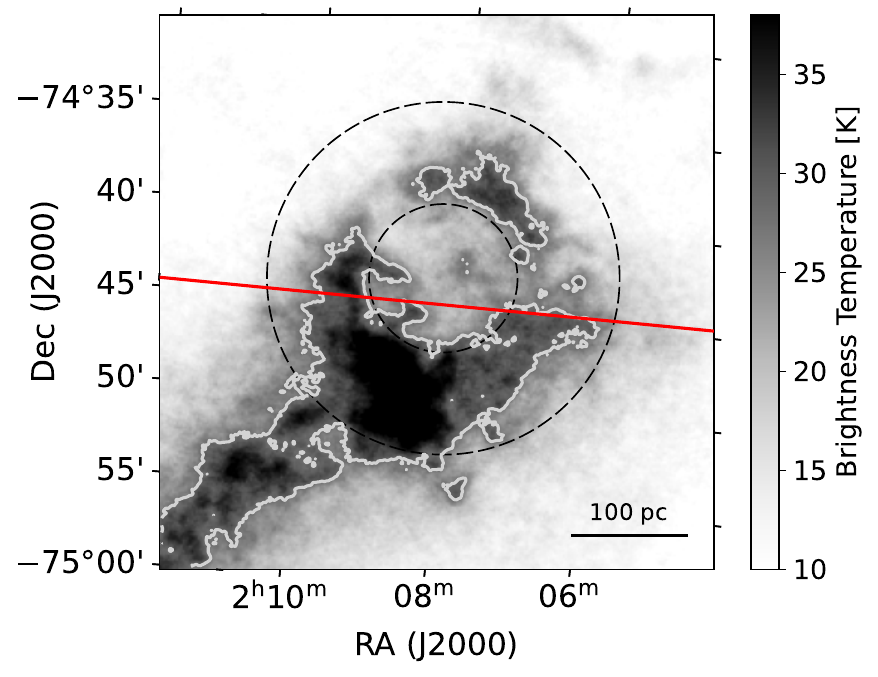}
    \caption{
    Mean brightness temperature map at 30\arcsec\,resolution of the 256$\times$256 pixel field centered on Shell-91, computed over the H{\sc i} emission velocity range of $170-175$\,\kms.
    A grey contour level of 28\,K is used to trace the shell structures.
    The black dashed circles denote the inner and outer boundaries of the shell, defined in Section~\ref{sec:3.1}.
    The red line traces the constant declination $-74^\circ 44^\prime 34^{\prime\prime}$ used to compute the RA-velocity diagram shown in Figure~\ref{fig:PVdiagram_with_G1234_contour}.
    A scale bar representing 100\,pc is shown in the lower-right panel.
    }
    \label{fig:peak_intensity.png}
\end{figure}
\begin{figure}
    \includegraphics[width=\columnwidth]{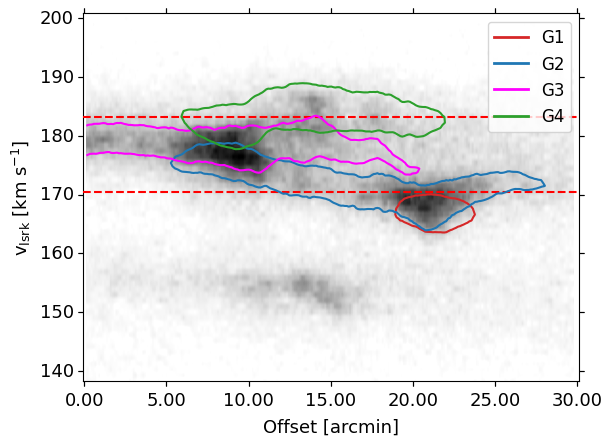}
    \caption{
    An RA-velocity diagram along declination $-74^\circ 44^\prime 34^{\prime\prime}$ of H{\sc i} Shell-91.
    The brightness temperature ranges from 5\,K to 55\,K. 
    The coloured solid lines show the contours of the contribution to the brightness temperature at 12\,K of the decomposed Gaussians G$_1$ to G$_4$, obtained with {\tt ROHSA}. Two dashed red lines indicate the approaching and receding velocities of Shell-91, 170 and 183\,\kms, respectively. 
    }
    \label{fig:PVdiagram_with_G1234_contour}
\end{figure}

To analyse Shell-91, we extracted a $256\times256$-pixel cube centered on the shell's location.  
Figure~\ref{fig:peak_intensity.png} shows the mean brightness temperature map of this sub-region, computed over the H{\sc i} emission velocity range $170-175$\,\kms. 
In this velocity interval, a central cavity is prominent, surrounded by enhanced brightness temperature at the shell boundaries.
The shell appears to be connected with a larger-scale H{\sc i} structure, extending from the south-east to the north-west.

We visually inspected the data cube to understand where the shell is located in Position-Position-Velocity (PPV) space. 
To highlight the structure of the data, we show in Figure~\ref{fig:PVdiagram_with_G1234_contour} an RA-velocity diagram computed along constant declination $-74^\circ 44^\prime 34^{\prime\prime}$ (red line in Figure~\ref{fig:peak_intensity.png}), similar to that shown in \citet[][see their figure 13]{2003MNRAS.339..105M}.
At low velocities (151 and 156\,\kms), we observe emission that is not connected to the shell emission seen at higher velocities (165 and 190\,kms) in the RA-velocity diagram. We therefore assume that this low-velocity feature is an unassociated velocity component, which we hereafter refer to as ``foreground emission". 
However, importantly for the rest of the analysis, while this foreground emission is well separated in the RA-velocity diagram, it is spectrally blended with emission at higher velocities, around $v=165$\,\kms. 
This blending is illustrated in the gray shaded area of Figure~\ref{fig:HI_shell_mosaic_spectrum_foreground} that shows an example of nine neighboring spectra where both of these components are visible.
At $\mathit{v} >$ 165\,\kms\,, the shell structure becomes discernible. 
At $\mathit{v}$ = 171.5\,\kms, a prominent shell with a bright rim and a circular shape is visible, with an angular diameter of 8\arcmin\ ($\sim$130\,pc at a distance of 56\,kpc). 
At $\mathit{v}\approx$ 187\,\kms\, the rim disappears.
In Figure~\ref{fig:PVdiagram_with_G1234_contour}, a circular shape is visible, which is a signature of expansion, and we annotated the approaching and receding sides of the shell identified by \citet{2003MNRAS.339..105M} with two red dashed lines.

\citet{2003MNRAS.339..105M} originally defined the shell centre based on a position-velocity (PV) diagram rather than in the RA-Dec plane, which shows offsets compared to the morphological shell center that can be seen in Figure~\ref{fig:peak_intensity.png}. 
To redefine the shell structures -- centre, inner rims, and outer boundaries -- we used the mean brightness temperature map (Figure~\ref{fig:peak_intensity.png}) where the shell inner rims are clearly visible.
Using a constant contour level of 28\,K to trace the rim structures, we define the shell centre and inner boundaries at RA = $02^{\mathrm{h}}08^{\mathrm{m}}08\fs3,\ $
Dec = $-74^{\circ}43^{\prime}16''$.
In Figure~\ref{fig:peak_intensity.png}, the inner black circle shows the size of the cavity (with a diameter of 8\arcmin).
The larger black circle represents the outer boundaries of the shell, also visually defined based on the channel maps, with a thickness of 5.5\arcmin.

\begin{table}
    \centering
    \begin{tabular}{lc}
        \hline
        Property & Value \\
        \hline
        Name & H{\sc i} Shell-91 \\
        R.A. (J2000) & 02$^{\rm h}$08$^{\rm m}$08.3$^{\rm s}$ \\
        Dec. (J2000) & $-74^\circ43'16''$ \\
        $V_{\rm lsrk}$ (\kms) & 170--175 \\
        Expansion Velocity (\kms) & 6.5 \\
        Radius (arcmin) & 4.0 \\
        Radius (pc) & 65 \\
        Dynamical Age (Myr) & 6 \\
        $\log E$ (erg) & 48.6 \\
        \hline
    \end{tabular}
    \caption{Physical properties of H{\sc i} Shell-91.}
    \label{tab:shell_properties}
\end{table}

\subsection{Foreground subtraction}\label{sec:foreground_sub}
\begin{figure}
    \includegraphics[width=\columnwidth]{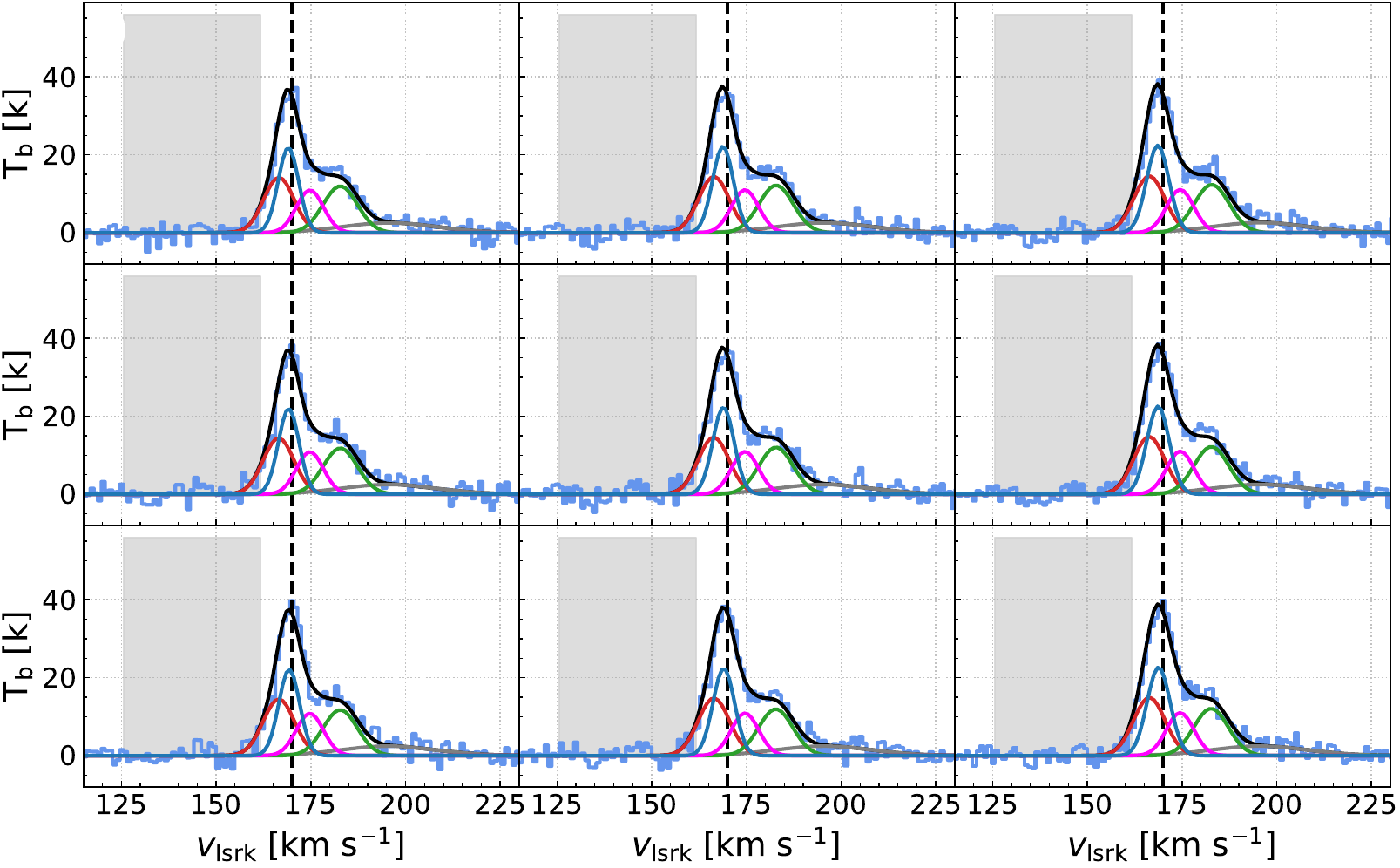}
    \caption{
    Example of nine neighboring spectra selected to highlight the Gaussian decomposition of the shell emission only using {\tt ROHSA}. 
    The solid blue lines display the H{\sc i} spectrum after foreground emission subtraction, while the solid black lines show the sum of the fitted Gaussian models. Individual Gaussian components are shown with coloured solid lines.} 
    The gray shaded area indicates the velocity range where the foreground emission dominates prior to subtraction. 
    The vertical dashed line indicates the CO peak velocity at $v_\mathrm{lsr}$ = 169.9\,\kms\, (\citealt{2006ApJ...643L.107M}).  
    \label{fig:HI_shell_mosaic_spectrum}
\end{figure}
Following \citet{2011ApJ...728..127D}, we used a Gaussian decomposition to subtract the foreground emission from the PPV cube.
We made use of the Regularized Optimization for Hyper-Spectral Analysis ({\tt ROHSA}; \citealt[hereafter M19]{marchal_2019}), which decomposes a data cube into multiple Gaussian components. 
%
%
{\tt ROHSA} fits the brightness temperature profiles of a PPV cube with a fixed number of Gaussian components for all sight-lines, and considering the spatial coherence between neighboring pixels. 
The multi-Gaussian model is given by
\begin{equation}
    \tilde{T}_b(v, \theta) = \sum_{n=1}^{N} a_n\exp\bigg(\frac{-(v-\mu_{n})^{2}}{2\sigma_{n}^{2}}\bigg),
\end{equation}
where $N$ is the number of Gaussian components provided by a user. $\theta$ represents the Gaussian parameters: amplitude $a_{n}$, velocity dispersion $\sigma_{n}$, and velocity $\mu_{n}$ of the {\it n}-th Gaussian component. 
The Gaussian parameters are found by minimizing a cost function described in \citetalias{marchal_2019}. 

In {\tt ROHSA}, there are three hyper-parameters ($\lambda_{a}, \lambda_{\mu}, \lambda_{\sigma}$) controlling the spatial coherency (i.e., smoothness) of the solution for each parameter map. 
We adopted values of $\lambda_{a}$, $\lambda_{\mu}$, and $\lambda_{\sigma}$ of 100, which balances the spatial coherence of neighboring pixels while maintaining the original beam resolution of the data. 
If this value is too low, the Gaussian parameter maps appear pixelated, whereas values that are too high lead to over-smoothed maps that deviate from the original beam resolution.

To model the foreground emission, we fitted the data cube in the low-velocity range only ($\mathit{v} \lesssim 160$\,\kms) with two Gaussians, and we found that these two components were sufficient to capture the emission at these velocities. 
%
These two Gaussians are shown with gray dashed lines, for the same set of spectra, in Figure~\ref{fig:HI_shell_mosaic_spectrum_foreground}. 
The column density maps of these components are shown in Figure~\ref{fig:filament}, which do not show any shell structures. 
We subtracted them from the original data cube, and the same set of spectra after foreground subtraction is shown in Figure~\ref{fig:HI_shell_mosaic_spectrum}. 
In the gray shaded area, the foreground emission has been effectively subtracted. 

\subsection{Gaussian decomposition of Shell-91}\label{rohsa}
\subsubsection{Exploring the parameter space}
We explored the number of Gaussians $N$ from 3 to 6 to decompose the signal of the shell ($160 \lesssim v \lesssim 200~\text{km s}^{-1}$), and chose the best model based on a global reduced chi-square criterion. 
For the noise terms in the chi-squared calculations, we used the standard deviation values measured from emission-free channels for each pixel. 
While increasing from $N$ = 3 to 5 improves the mean chi-squared values from 1.4 to 1.2, increasing to $N$ = 6 shows only a marginal improvement, with a mean chi-squared value remaining at 1.2. 
Thus, we chose $N$ = 5 to favour the simplest model and prevent overfitting of the data. 
The resulting mean, median, and standard deviations of the chi-squared values across the fields are 1.2, 1.2, and 0.2, respectively. 
We also made use of the fourth hyper-parameter, $\lambda'_{\sigma}$, available in {\tt ROHSA}, which controls the variance of each dispersion field. 
It is particularly useful when trying to perform phase separation, as its aim is to cluster the Gaussians along the dispersion axis of the parameter space. 
Specifically, we empirically chose a value of 100 for $\lambda'_{\sigma}$, as lower values prevent clustering, while higher values lead to an overly flat distribution in $\sigma$.
The impact of {\tt ROHSA} parameters on the cold H{\sc i} gas distributions is discussed in Appendix~\ref{app:uncertainties}, and it shows that the resulting map remains stable under different parameter choices. 
In Figure~\ref{fig:HI_shell_mosaic_spectrum}, we show an example of the fitted Gaussians for the same mosaic of nine neighboring spectra. 

\subsubsection{$\sigma-\mu$ diagram}
\begin{figure}
    \includegraphics[width=\columnwidth]{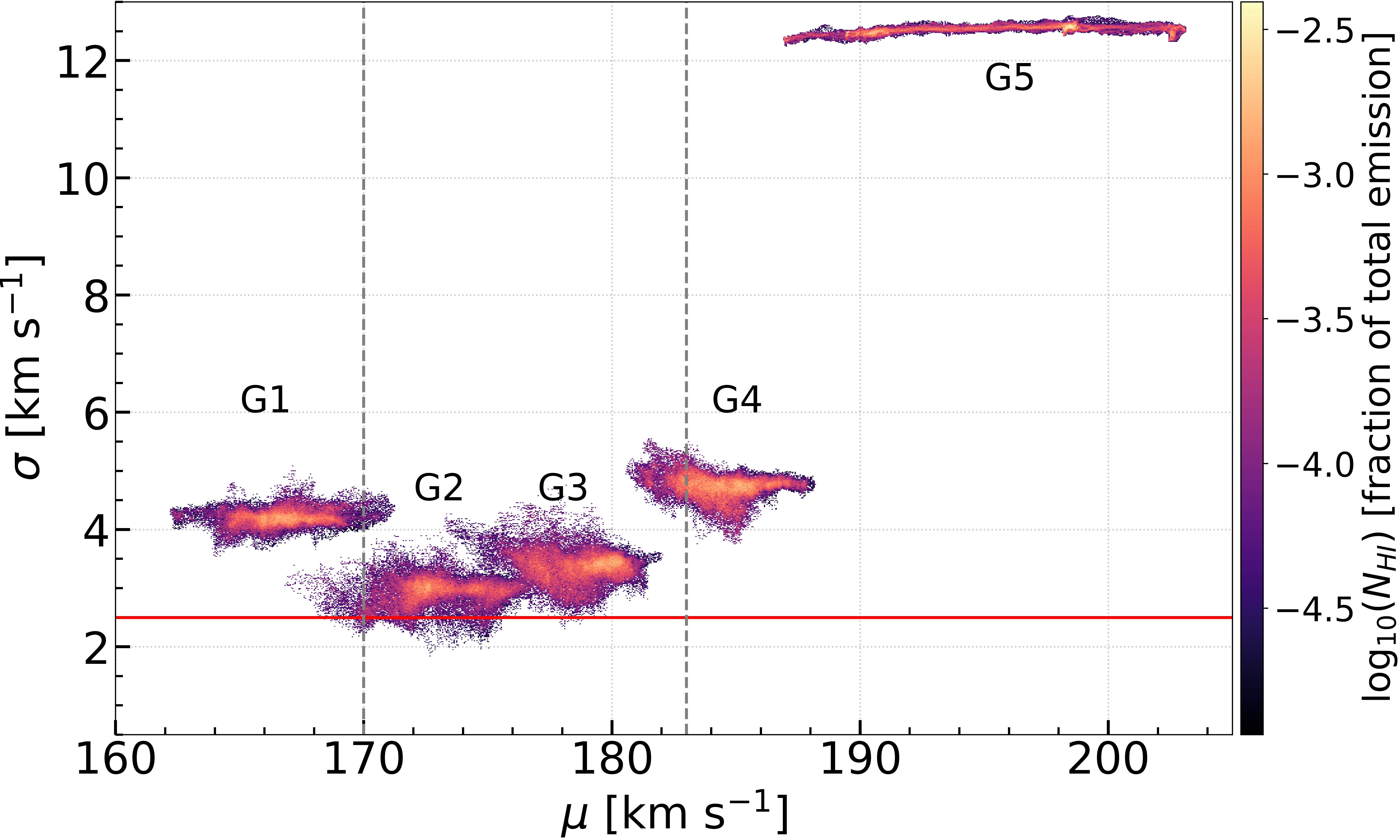}
    \caption{
    A 2D histogram of the $\sigma-\mu$ diagram for the decomposed H{\sc i} Shell-91, weighted by column densities. 
    The two vertical dashed gray lines indicate the approaching and receding velocities of H{\sc i} Shell-91, 170 and 183\,\kms, respectively.
    Gaussians are labeled from G$_1$ to G$_5$.
    The red horizontal line shows the cut used to extract the narrowest components.
    }
    \label{fig:sigma_mu_diagram}
\end{figure}
\begin{table}
\centering
\small
\begin{tabular}{c c c c c c} 
 \hline
 & $G_1$ & $G_2$ & $G_3$ & $G_4$ & $G_5$ \\ 
  [0.5ex] 
 \hline
$\langle \mu_n \rangle$ & 166.7 & 172.8 & 178.5 & 184.3 & 195.3 \\
$\langle \sigma_n \rangle$ & 4.2 & 2.9 & 3.4 & 4.7 & 12.5 \\
 \hline
\end{tabular}
\caption{H{\sc i} Shell-91 properties: Column-density-weighted mean centroid velocities ($\mu_n$) and velocity dispersions ($\sigma_n$) in units of \kms\, for the five Gaussian components decomposed by {\tt ROHSA}.
}
\label{table:1}
\end{table}
In Figure~\ref{fig:sigma_mu_diagram}, we show the column density-weighted 2D histograms of $\sigma$ and $\mu$.
The five clusters, corresponding to the five Gaussian components, are well-separated in $\sigma-\mu$ space, with velocity dispersions ($\sigma$) ranging from $\sim$2 to 12\,\kms, and the velocities ($\mu$) ranging from $\sim$160 to 200\,\kms. 
The column-density-weighted mean centroid velocities ($\mu_n$) and velocity dispersions ($\sigma_n$) of the five Gaussian components are tabulated in Table~\ref{table:1}.
The velocities of the approaching and receding sides, 170 and 183\,\kms\,, respectively, are annotated by two vertical gray dashed lines in Figure~\ref{fig:sigma_mu_diagram}. 
We observe that the narrower Gaussian components G$_2$ and G$_3$ are located at inner velocities, while broader components G$_1$, G$_4$, and G$_5$ are found at outer velocities. 
They form a ``V" shape in this diagram. 
On average, G$_2$ exhibits lower velocity dispersions than G$_3$ and also contains the narrowest dispersion values observed across the entire parameter space.
To highlight this, a horizontal red line marks the $\sigma = 2.5$\,km\,s$^{-1}$ threshold as the criterion for cold H{\sc i}, indicating that a part of the G$_2$ component falls below this value.

Finally, to place the $\sigma-\mu$ diagram in context with the RA–velocity diagram, we overplot in Figure~\ref{fig:PVdiagram_with_G1234_contour} the 12\,K brightness temperature contours for each Gaussian component from G$_1$ to G$_4$. The 12\,K brightness temperature level is arbitrary and is chosen for visualization, as the G$_1$ to G$_4$ components exhibit different brightness temperature ranges. G$_5$ is not included, as its large velocity dispersion ($\langle \sigma_n \rangle$ = 12.5 \kms) suggests it likely traces surrounding diffuse warm gas, or it could be an artifact of the baseline fitting. Crucially, unlike the G$_1$--$G_4$ components, G$_5$ shows no clear indication of an association with the shell (see Figure~\ref{fig:mosaic_field_map} for the column density maps).

\subsubsection{Column density maps}
\begin{figure*}
    \centering
    \includegraphics[width=0.95\textwidth]{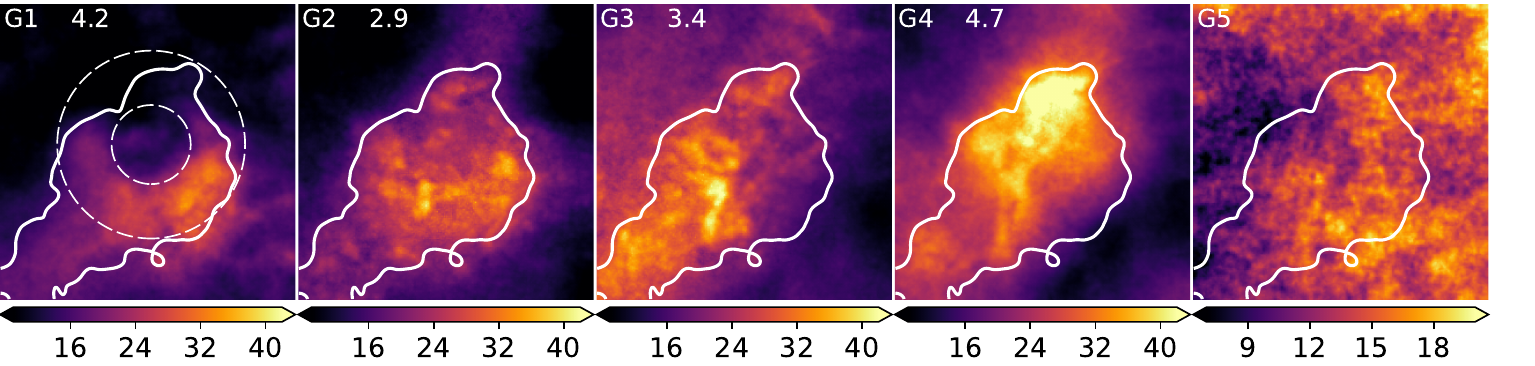}
    \includegraphics[width=0.95\textwidth]{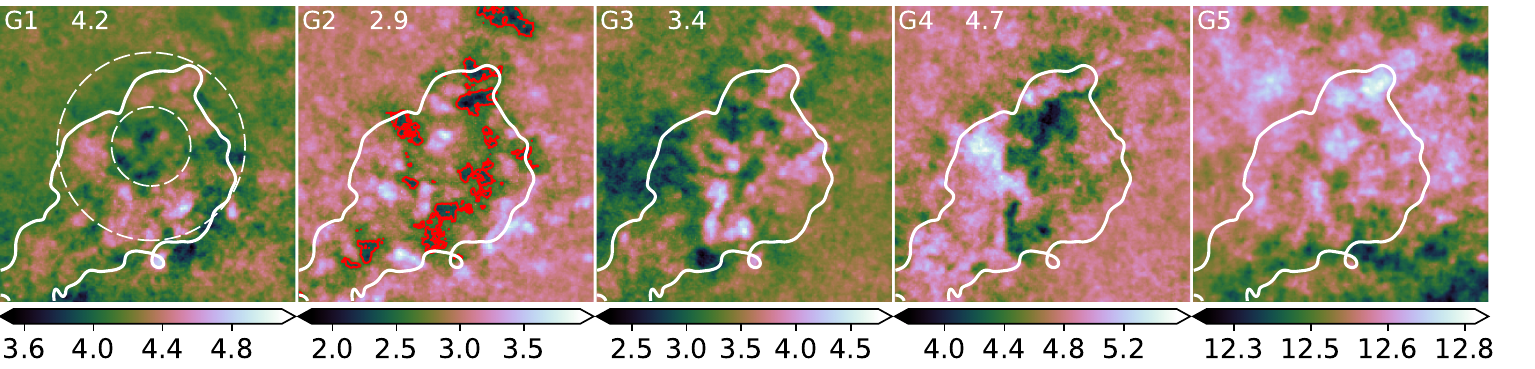}
    \includegraphics[width=0.95\textwidth]{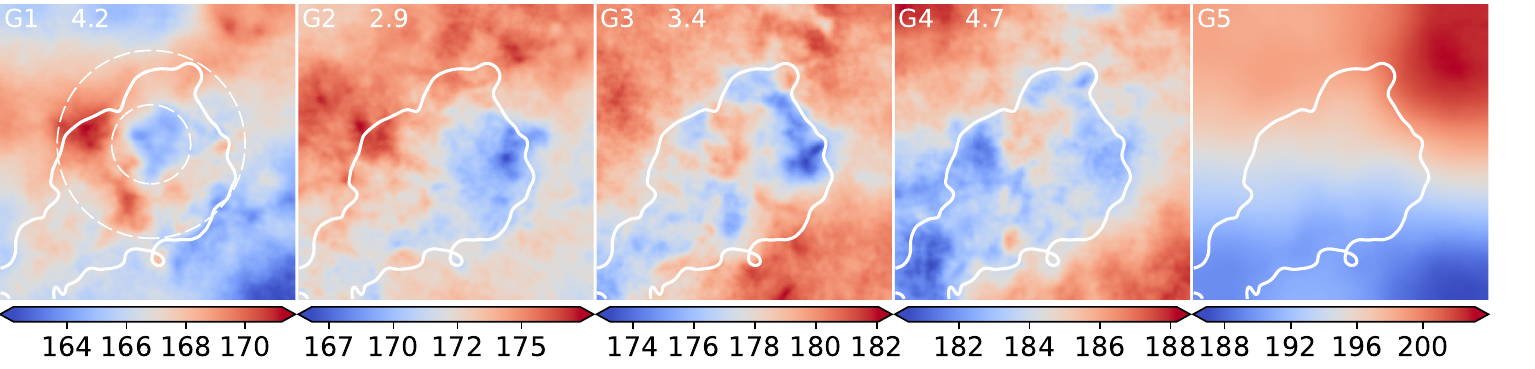}
    \caption{
    Column density (in units of 10$^{19}$\,cm$^{-2}$; top), derived under the optically thin assumption, velocity dispersions (in units of \kms; middle), and centroid velocity maps (in units of \kms; bottom) of individual Gaussian components of the H{\sc i} Shell-91, sorted by increasing centroid velocity. 
    The white solid contour indicates the total column density at a level of 10$^{21}$\,cm$^{-2}$. 
    These contours are derived from the foreground-subtracted Position-Position-Velocity (PPV) cube. 
    The two white dashed circles in the leftmost panels indicate the inner and outer boundaries of Shell-91, obtained from the H{\sc i} channel maps in Section~\ref{sec:3.1}.
    From left to right, the panels show G$_1$-G$_5$ components. 
    %
    %
    The red contours in the middle G$_2$ panel highlight cold gas components with velocity dispersions $\sigma < 2.5\,\kms$.
    The number on each top panel indicates the column-density-weighted velocity dispersion in units of \kms.}
    \label{fig:mosaic_field_map}
\end{figure*}
\begin{figure*}
    \centering
    \includegraphics[width=\linewidth]{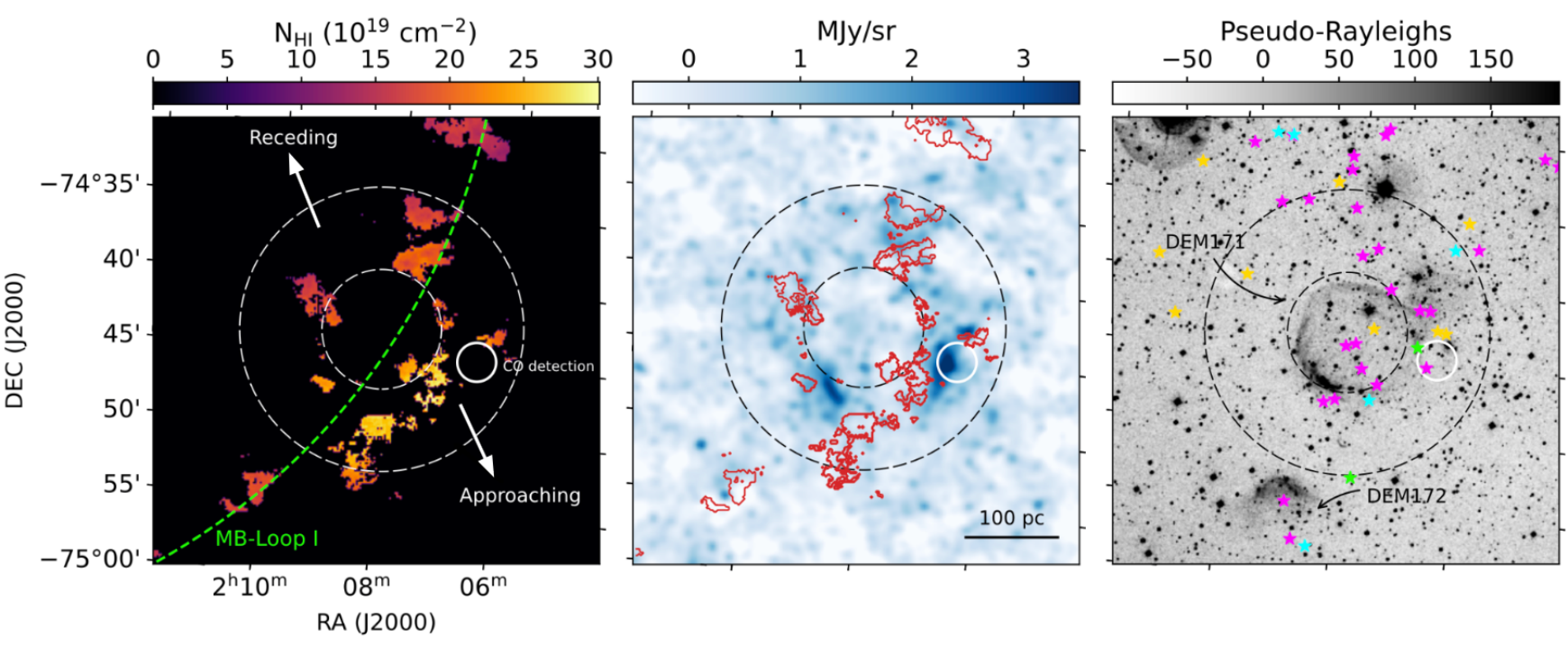}
    \caption{
    Left: Column density of cold H{\sc i} gas with $\sigma < 2.5$\,km\,s$^{-1}$ extracted from the Gaussian component G$_2$.
    Two white dashed circles denote the inner and outer boundaries of Shell-91, obtained from the H{\sc i} channel maps described in Section~\ref{sec:3.1}.
    The solid white circle shows the location of the CO cloud (region E) cataloged by \citet{2006ApJ...643L.107M}.
    Middle: Intensity at 250$\micron$ from \textit{Herschel} SPIRE in unit of MJy\,sr$^{-1}$. 
    The red contours indicate the locations of the cold clouds ($\sigma < 2.5\,\kms$) found with {\tt ROHSA}.
    Right: SuperCOSMOS H$\alpha$ image of Shell-91/DEM171 identified by \citet{1986MNRAS.223..317M}. Intensity values are not accurate due to calibration uncertainties. 
    The different color markers represent the star clusters (magenta) from \citet{2020AJ....159...82B}, and the YSOs (lime), faint YSOs (yellow), and HAeBe (cyan) objects from \citet{2014ApJ...785..162C}. 
    A scale bar representing 100\,pc is shown in the middle panel.} 
    \label{fig:three_plots}
\end{figure*}

Figure~\ref{fig:mosaic_field_map} shows the column density maps of the five Gaussian components (top row), ordered by increasing mean centroid velocity. 
These column density maps are derived under the assumption of an optically thin medium.
The corresponding velocity dispersion and centroid velocity maps are shown in the second and third rows from the top.
%
%
The dispersion field of each Gaussian reveals significant fluctuations, reflecting variations in temperature and/or turbulent broadening of the gas.
We observe that the broader components G$_1$ and G$_4$ exhibit an asymmetry in their column density distributions, with most of the mass in G$_1$ and G$_4$ located on the south-western and north-eastern sides of the shell, respectively.

The component G$_2$ is of particular interest, as it has the lowest mean velocity dispersion among all components. 
In particular, the dispersion field of G$_2$ shows that regions with velocity dispersion below $2.5$\,\kms -- highlighted by the red contours (corresponding to the red line in Figure~\ref{fig:sigma_mu_diagram}) -- are spatially clustered on small scales. 
We also note, from the velocity field of G$_2$, that the cold/narrow clouds located on the south-western side (i.e., the approaching side) are moving toward lower velocities than those on the opposite side (north-eastern or receding side), whose velocities tend to be closer to the central velocity of the shell.
This asymmetry can also be seen in the RA-velocity diagram, where the blue (G$_2$) and red (G$_1$) contours coincide on the western side.

For a better view of the spatial distribution of these narrow features, we created a column density map including only pixels with velocity dispersions below 2.5 km s$^{-1}$ extracted from the G$_2$ components. The resulting map is shown in Figure~\ref{fig:three_plots}.
We overplotted, with white dashed circles, the inner and outer boundaries of the shell obtained from the mean brightness-temperature map computed over the narrow velocity ranges of Shell-91 described in Section~\ref{sec:3.1}.
Interestingly, these narrowest clouds are well located within the shell's 
boundaries, with the exception of two islands of clouds located in the south-eastern part and north-western part of the field. 
The spatial distribution of these clouds appears to delineate a larger arc-like structure (hereafter referred to as MB-Loop~I; see Section~\ref{sec:loop}), highlighted with a green dashed line. 
The south-west/north-east asymmetry seen in G$_1$ and G$_4$, which drives the shell's expansion (annotated with white arrows here), appears roughly perpendicular to the line traced by those cold clouds (green dashed line). 
This geometry will be discussed further in Section~\ref{sec:discussion} in the context of the larger-scale structure of the region.
Finally, we note that the column densities of cold gas range from about 15 to 30\,$\times$10$^{19}$\,cm$^{-2}$. 
Assuming an optically thin medium, we find that in this region, the total column densities are about a factor of 10 higher than those of the cold clouds. 
We note that if these cold clouds have high optical depth, the actual column densities of these cold clouds would increase upon an opacity correction.

\section{Star formation tracers within Shell-91}
Having characterized the cold H{\sc i} structure and kinematics of Shell-91, we now investigate its associated molecular gas content, physical properties, and surrounding ionized gas to assess the relationship between the shell and tracers of star formation. 

\subsection{Molecular content}
We have found only one observation of ${}^{12}\text{CO} \, (1-0)$ emission \citep[region E in][]{2006ApJ...643L.107M}, directly tracing the molecular gas content within the shell.
We show in Figure~\ref{fig:three_plots} (left) the location and beam size of this observation, as well as the peak velocity of the observed CO line in Figure~\ref{fig:HI_shell_mosaic_spectrum}.
We find that the CO peak velocity at $v=169.9$\,\kms\, coincides with a bright component in G$_2$ with a velocity dispersion around 3\,\kms.
In projection, this CO emission is located within the shell and appears to be surrounded by narrow-linewidth clouds. 
It is important to note that the ${}^{12}\text{CO}\,(1-0)$ observations obtained by \citet{2006ApJ...643L.107M} were targeted toward regions of bright 100\,$\mu$m emission detected by IRAS. 
This is because far–infrared dust emission provides an indirect tracer of the total gas column density and therefore of regions where molecular gas may form \citep[e.g.,][]{Boulanger:1996, PlanckCollaboration:2011}.
However, there is currently no comprehensive mapping of direct molecular gas tracers in this region. 
While this single pointing necessarily biases our view of the molecular content of the shell, it nevertheless indicates (together with our multiphase analysis of the H{\sc i} gas) the presence of a mixture of cold atomic and molecular gas within Shell-91. 

Because targeted observations of co-spatial of bright H\,{\sc i} ($N_{\text{HI}} \gtrsim 10^{21} \text{ cm}^{-2}$) and 100\,$\mu$m emission have led to detections of CO molecules (\citealt{2006ApJ...643L.107M}), it is informative to place the cold clouds revealed by our decomposition in the context of a fully sampled map tracing dust emission in the far-infrared. 
\textit{Herschel} SPIRE 
250\,$\mu$m predominantly arises from large dust grains in thermal equilibrium with the interstellar radiation field and is approximately proportional to the dust column density in the optically thin regime \citep[e.g.,][]{Boulanger:1996, 2012MNRAS.422.2291P}. 
Because dust and gas are generally well mixed in the ISM, enhanced 250\,$\mu$m surface brightness typically corresponds to regions of elevated total hydrogen column density (e.g., \citealt{2015ApJ...799...96G}). 
However, the dust-to-gas ratio (DGR) varies across the Magellanic Clouds and has very low values in the SMC tail regions (see; \citealt{2009ApJ...690L..76G, 2012ApJ...745..173W}).
In such low-DGR and low-metallicity environments, more H{\sc i} column densities are required for the formation of H$_2$ and CO molecules through self-shielding and dust shielding against dissociating ultraviolet radiation \citep{Wolfire:2010,Sternberg:2014, 2009ApJ...693..216K}.
%
%
Consequently, far–infrared maps provide a useful way to identify locations where molecular gas may be present, even in the absence of fully sampled CO observations \citep[e.g.,][]{PlanckCollaboration:2011}.
Specifically, we made use of the Herschel product at 250\,$\mu$m, which is part of the HERschel Inventory of The Agents of Galaxy Evolution (HERITAGE; \citealt{2013AJ....146...62M, 2014ApJ...797...85G}).
The SPIRE 250\,$\mu$m map has an 18.2\arcsec resolution, and the Milky Way foreground emission has been subtracted and processed by \citet{2014ApJ...797...85G}.

Figure~\ref{fig:three_plots} (middle panel) shows the intensity at 250\,$\micron$ in units of MJy\,sr$^{-1}$. Annotations are as in the left panel, and red contours indicate the locations of the cold clouds with $\sigma < 2.5\,\kms$.
In addition to a homogeneous background of small-scale fluctuations characteristic of the cosmic infrared background (CIB) \citep{Auclair:2024}, we observe a clear enhancement of FIR emission where CO was detected by \citet{2006ApJ...643L.107M}, and interestingly we find a second region of similar FIR intensity at $\mathrm{RA} = 02^{\mathrm{h}}08^{\mathrm{m}}37^{\mathrm{s}}$, $\mathrm{Dec} = -74^{\circ}47^{\prime}29''$.
Similar to the location of the CO detection, this FIR emission is surrounded by cold H{\sc i} clouds that were revealed with {\tt ROHSA}. 
This suggests that the CNM gas may form an external layer around the CO clouds rather than being fully co-spatial with them. 
A detailed characterization of the relationship between cold H\,{\sc i} gas and the molecular phase is beyond the scope of this paper and would require new CO observations.

\subsection{Dynamical age and Energy}
To understand the origin and formation mechanism of the shell, we estimate its dynamical age and the energy required to create this expanding shell, with our refined shell size ($R_s$) and expansion velocity ($V_\mathrm{exp}$). 
For the dynamical age in Myr ($t_\mathrm{dyn}$), we adopt the stellar-wind-blown bubble model by \citet{1977ApJ...218..377W}:
\begin{equation}
    t_\mathrm{dyn} = \frac{3}{5} \left( \frac{R_s}{V_\mathrm{exp}} \right),
\end{equation}
where $R_s$ is the shell radius in pc, and $V_\mathrm{exp}$ is the expansion velocity in \kms.
Here, we adopt a shell radius of $R_s = 65$\,pc and derive an expansion velocity of $V_\mathrm{exp} = 6.5$\,\kms\, calculated as half the velocity difference between the receding (183\,\kms) and approaching (170\,\kms) velocities (($V_\mathrm{receding}$ - $V_\mathrm{approaching}$)/2; see Figure~\ref{fig:PVdiagram_with_G1234_contour}).
Based on the measured shell radius and expansion velocity, we obtain a dynamical age of 6\,Myr.
%
This value is slightly lower than the $9\pm2$\,Myr reported by \citet{2003MNRAS.339..105M}.
The primary difference is due to our adoption of a smaller shell size and a closer distance to the Magellanic Bridge (e.g., 56\,kpc in our study, assuming an intermediate LMC--SMC distance, compared to the 60\,kpc in \citealt{2003MNRAS.339..105M}).

To estimate the energy required to create the shell, we calculated the luminosity injected into the ambient medium using the stellar-wind model by \citet{1977ApJ...218..377W}:
\begin{equation}
    L_s = 1.5 \times 10^5 \left( \frac{R_s}{100\,\mathrm{pc}} \right)^5 \left( \frac{t_\mathrm{dyn}}{10^6\,\mathrm{yr}} \right)^{-3} \left( \frac{n_0}{1\,\mathrm{cm}^{-3}} \right) L_\odot,
\end{equation}
where n$_0$ is the ambient density and $L_\odot = 3.9 \times 10^{33}$\,\text{erg\,s$^{-1}$} is the solar luminosity. 
The total energy ($E$) in erg is derived by multiplying this luminosity by the dynamical age of the shell in seconds ($E = L_s \times t_\mathrm{dyn}$).
Adopting an ambient density of $n_0 = 0.06$\,\text{cm$^{-3}$} (\citealt{2003MNRAS.339..105M}), we derive a total energy of $\log (E / \mathrm{erg}) \approx 48.6$.
This value is consistent with $\log (E / \mathrm{erg}) \approx 48.7$ reported by \citet{2003MNRAS.339..105M}.

We note that the derived physical properties carry uncertainties. The shell size and the expansion velocity are measured directly from our high-resolution spatial and spectral data, so the dynamical age is relatively well constrained. However, the ambient density of $n_0 = 0.06$\,\text{cm$^{-3}$} used both in \citealt{2003MNRAS.339..105M} and in our study is poorly constrained. This value relies on assumptions regarding the line-of-sight depth, which was used to derive the volume of the Bridge, introducing uncertainties up to a factor of a few. Consequently, the energy required to form the shell should be taken as an order-of-magnitude estimate.
We tabulate the physical properties of the H{\sc i} shell in Table~\ref{tab:shell_properties}, including its location, size, expansion velocity, dynamical age, and the total energy required to drive its expansion.

\subsection{Ionized phase and ionizing mechanism}
To investigate the spatial distribution of the ionized phase of Shell-91, we made use of the SuperCOSMOS H$\alpha$ Survey (SHS; \citealt{2005MNRAS.362..689P}). 
We performed a background subtraction of the original data using the {\sc photutils.background} package. 
Additionally, we have applied a 3 $\times$3 pixel median filter to the data to enhance the diffuse emission. 
The right panel in Figure~\ref{fig:three_plots} shows the resulting H$\alpha$ intensity in units of pseudo-Rayleighs. 
As mentioned in \citet{2007PASA...24...69M}, there are significant challenges in obtaining accurate calibration. 
Therefore, we note that the intensity values shown here should be taken with caution.
The ionized shell structure of DEM171 is clearly visible, as well as the DEM172 feature to the south.
We find that the H$\alpha$ shell DEM171 aligns well with the inner boundary of the H{\sc i} Shell-91, as indicated by the inner white circle obtained from the H{\sc i} channel maps described in Section~\ref{sec:3.1}. This suggests that the shell interior is filled with ionized gas, while the shell boundaries consist of swept-up neutral gas material, mixed with molecular gas.

The ionizing mechanism for DEM171 remains unclear. 
Previous studies have suggested that this H$\alpha$ shell could be ionized by stellar winds or supernovae, and that its expansion could be driven by a Wolf-Rayet (WR) star \citep{1986MNRAS.223..317M, 2001MNRAS.326..539G, 2007PASA...24...69M}.
The estimated energy of $\log (E / \mathrm{erg}) \approx 48.6$ is much lower than the typical energy injected by a single supernova ($\sim 10^{51}$\,\text{erg}).
Even assuming an ambient density higher by an order of magnitude ($n_0 = 0.6\,\mathrm{cm^{-3}}$), the resulting total energy of $\log (E / \mathrm{erg}) = 49.6$ ($\approx4\times10^{49}$\,erg) remains below $10^{51}$\,erg.
For comparison, the total energy injected by the WR wind during the WR phase has been estimated to be $7\times10^{50}$\,erg (\citealt{1981MNRAS.196..101B}), which is much larger than the estimated energy for the DEM171 shell.
This suggests that the SNe or WR stellar winds are unlikely to be the primary driving mechanisms for the shell. 
Instead, the shell could be powered by stellar winds from lower-mass or late-type massive stars.
%

To identify potential ionizing sources, we investigate the spatial distributions of the surrounding stellar populations.
The magenta stars in Figure~\ref{fig:three_plots} (right) indicate the positions of star clusters compiled by \citet{2020AJ....159...82B}, and the lime, yellow, and cyan stars show the young stellar objects (YSOs), faint YSOs, and Herbig Ae/Be stars (HAeBe or intermediate-mass YSO) from \citet{2014ApJ...785..162C}, respectively. 
The star clusters (gray) near the shell centre are potential ionizing sources for the DEM171 feature, and clusters near the DEM172 region are also candidates for its ionization. 
Interestingly, several YSOs (ages $\lesssim 1\text{--}2$\,Myr; \citealt{2014ApJ...785..162C}) are found within the shell structures, with some located near the molecular cloud Region E (in projection) at the shell boundary. 
Because these YSOs ($\lesssim 1\text{--}2$\,Myr) are younger than the dynamical age of the shell (6\,Myr), they are unlikely drivers of ionization; instead, they represent candidates for recent, potentially triggered star formation along the shell boundaries.

\section{Large-scale context of the cold neutral phase}
While the previous sections focused on the shell and its immediate environment, we now place Shell-91 in the broader context of the cold gas distribution across the Magellanic Bridge.
\label{sec:FFT}

\begin{figure*}
	\includegraphics[width=0.9\linewidth]{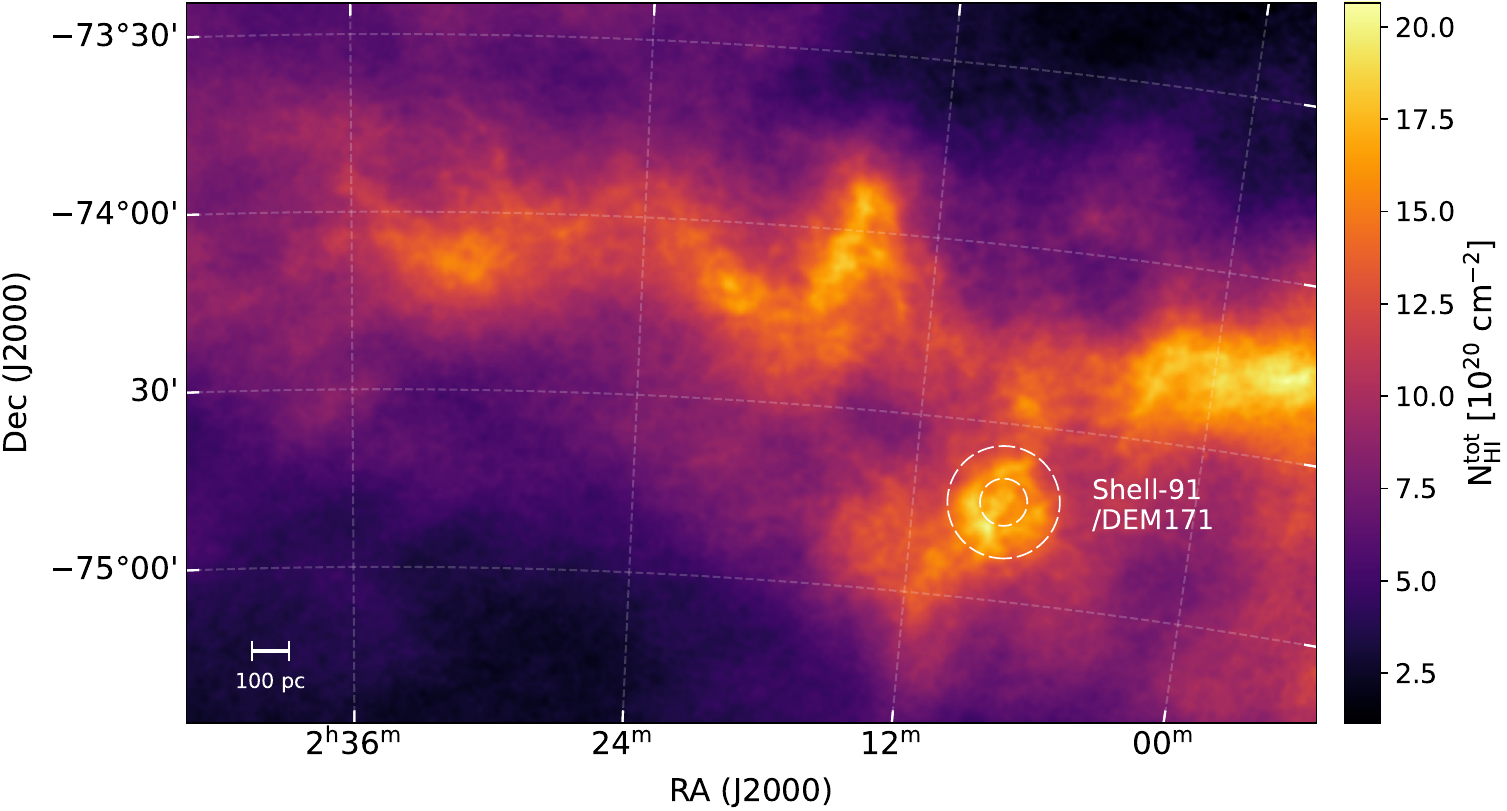}
    \includegraphics[width=0.9\linewidth]{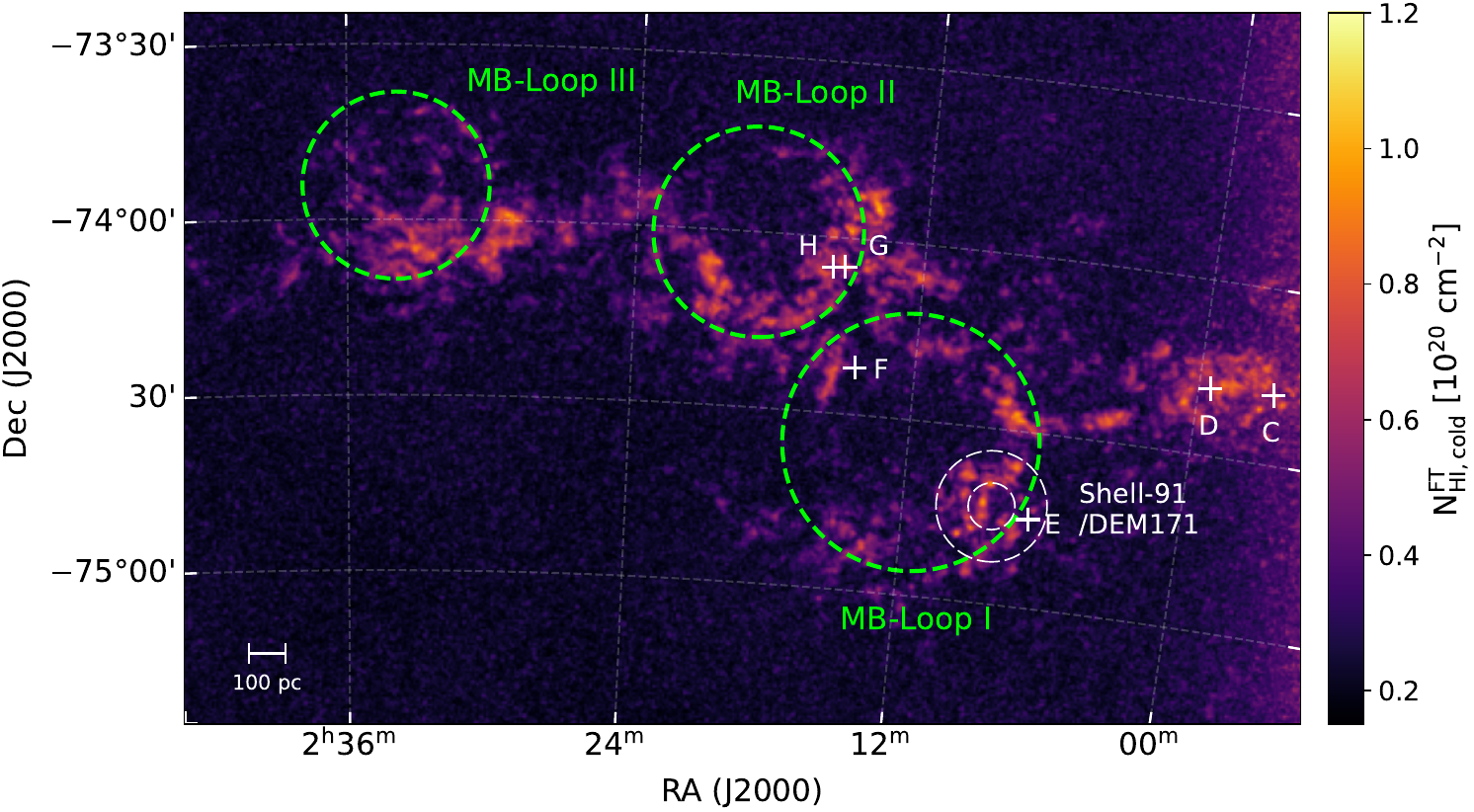}
    \caption{
    Top: Total H{\sc i} column density map of the western side of the Magellanic Bridge, computed in the optically thin limit. 
    Bottom: Lower limit of the cold gas column density map of the western side of the Magellanic Bridge inferred from the FT method. 
    The dashed green circles denote the location of the three large-scale loops identified in the FT map.
    Two white dashed circles denote the inner and outer boundaries of Shell-91, defined in Section~\ref{sec:3.1}.
    The white crosses indicate the positions of molecular clouds (regions C to H) traced by ${}^{12}\text{CO} \, (1-0)$ emission observed with NANTEN with a beam size of 2.6\arcmin.
    A scale reference of 100\,pc is annotated in the lower left part of the maps.}
    
    \label{fig:FFT}
\end{figure*}
\subsection{FT method and cold gas map}
The GASKAP–HI observations used here cover a region of the Magellanic Bridge that is significantly larger than the extent of Shell-91. 
This allows us to investigate the large-scale environment within which the shell is embedded.
While applying {\tt ROHSA} to such large regions is challenging and beyond the scope of this paper, we instead adopt the methodology developed by \citet[][hereafter M24]{2024ApJ...961..161M}, which provides a first-order view (i.e., a lower limit of the cold gas column density) of the spatial distribution of cold gas in 21\,cm data cubes.
Specifically, this method applies a Fourier Transform (FT) to the H{\sc i} emission profile along each line of sight, transforming the brightness temperature information from the velocity space to the Fourier domain defined by the wavenumber $k_v$. In this domain, cold gas with narrow linewidth is captured at high $k_v$, while warm gas with broad velocity dispersions only contributes at low $k_v$. To isolate the cold gas, we used a threshold of $k_{\lim}=0.12$$(\kms)^{-1}$, which ensures that any Gaussian feature with a velocity dispersion of 3\,\kms (or higher) is suppressed at the level of 95\% in the total integrated intensity of the line. 
An important caveat of the FT method is its sensitivity to noise in low signal-to-noise regions, as noise tends to introduce spectral structure at high $k_v$. 
To reduce these effects, we performed the FT on a data cube convolved to a 1\arcmin\,beam (lowering the resolution by a factor of two) and resampled at the Nyquist rate, enhancing the signal-to-noise ratio by a factor of $\sqrt{2}$. 
The top panel in Figure~\ref{fig:FFT} shows the total H{\sc i} column density of the field computed in the optically thin limit, and the bottom panel shows the lower limit of the cold gas column density. 
In Figure~\ref{fig:FFT}, we have annotated with white crosses the locations of all molecular clouds (Regions C to H; ${}^{12}\text{CO} \, (1-0)$) cataloged by \citet{2006ApJ...643L.107M}. Note that no CO observations were conducted beyond $02^{\mathrm{h}}20^{\mathrm{m}}$. 
As for Region~E, these targeted observations provide only limited information on the locations where molecular gas is present in the Bridge. Nevertheless, we find that these molecular clouds are generally located in regions of enhanced cold gas column density, as revealed by the FT method.
These high cold H{\sc i} column density regions do not necessarily represent physically coherent structures, but rather the regions of enhanced cold H{\sc i} gas along the line of sight.

\subsection{Loop structures and morphology}\label{sec:loop}
As shown in Figure~\ref{fig:FFT}, cold H{\sc i} gas appears to be structured over a wide range of scales, forming loops and arcs all along this part of the Magellanic Bridge. 
Specifically, the FT map reveals three large-scale open loops that we have annotated as MB-Loop~I, II, and III, where MB refers to the Magellanic Bridge.
MB-Loop~II and III are preferentially open toward the north, while MB-Loop~I seems to be open on its eastern side.
The white dashed circle indicates the location and approximate size of Shell-91.
There seems to be a relationship between the large-scale MB-Loop~I and the small-scale Shell-91, which is located along the loop in projection. 
Furthermore, it is clear that the two islands of cold H{\sc i} gas located outside the shell boundaries, as described in Figure~\ref{fig:three_plots} (left), are part of MB-Loop~I, that we annotated as well on Figure~\ref{fig:three_plots} using the same dashed green line.
%
Inspecting the PPV cube visually, we find that MB-Loop~I is composed of superimposed structures spanning various velocity ranges; however, its southern base displays arc-like structures extending from north to south-east at around 175\,\kms, coinciding with the velocity of Shell-91. At this specific velocity, Shell-91 appears kinematically linked to this arc along the loop.


\section{Discussion} \label{sec:discussion}
\subsection{Small-scale: cold gas within Shell-91}

%
Simulations have shown that converging flows can promote the atomic-to-molecular phase transition through gas accumulation, condensation, thermal instability, and radiative cooling (e.g., \citealt{2009ApJ...704..161I, 2012ApJ...759...35I, 2015A&A...580A..49I}). 
Expanding shells or bubbles created and powered by stellar feedback sweep up and compress the surrounding medium, leading to gas accumulation at their boundaries and creating conditions analogous to converging flows. %
This compression can trigger thermal instability and radiative cooling, leading to the formation of cold H{\sc i} and molecular gas at shell boundaries.

As mentioned in Section~\ref{sec:intro}, observational evidence supporting this scenario has been found in the Milky Way and the Magellanic Clouds supershells. A few studies have reported cold H{\sc i} associated with shells (\citealt{2000A&A...354..787M, 2003ApJ...594..833M, 2011ApJ...728..127D, 2012MNRAS.421.3159M}). 
Additionally, there is evidence of molecular cloud formation at the boundaries of shells in the LMC and the Milky Way. 
For example, in the LMC, \citet{2013ApJ...763...56D} estimated that at least $4-11\%$ of the total molecular cloud mass likely formed as a result of large-scale stellar feedback. In the Milky Way, \citet{2011ApJ...728..127D} showed a statistical enhancement of molecular gas within the volumes occupied by two supershells (GSH 287+04−17 and GSH 277+00+36) compared to local background regions. 

In agreement with previous findings, Figure~\ref{fig:three_plots} shows that Shell-91 contains cold H{\sc i} gas clumps ($\sigma <$ 2.5\,\kms). 
A linewidth of 2.5\,\kms\, corresponds to a maximum kinetic temperature of $\sim$750\,K; though the true kinetic temperature should be lower if turbulence contributes to the observed linewidth. 
The other H{\sc i} components of Shell-91 (G$_1$, G$_3$, and G$_4$), decomposed by {\tt ROHSA} in Section~\ref{rohsa}, exhibit broader mean velocity dispersions of 3-5\,\kms (see Figure~\ref{fig:mosaic_field_map}).
One molecular cloud with an estimated mass of 10$^3$\,\Msun\, (\citealt{2006ApJ...643L.107M}) traced by CO observed with the NANTEN telescope is also detected within Shell-91.
The presence of the molecular cloud supports the idea that this shell provides an environment that harbors a cold and dense gas phase. 
Our work shows that this shell wall is made of a mixture of cold atomic and molecular gas, surrounded by gas with slightly higher velocity dispersion (i.e., more turbulent cold gas and/or warmer gas), while the central cavity is ionized.
However, the complex environment of a stellar-feedback-driven shell makes it difficult to determine the exact origin of this molecular cloud—whether it is a pre-existing parent cloud or formed in situ from cold atomic gas.
In addition, the origin of the cold H{\sc i} gas within shell boundaries remains uncertain. 
Two cold gas clumps are present in the outer region of the H{\sc i} Shell-91 analyzed here (Figure~\ref{fig:three_plots}), appearing to be located along the large-scale loop structure.
As Shell-91 is embedded within the large-scale arc structure extending from south to north (MB-Loop~I), this suggests that Shell-91 might have formed out of this structure, and some cold gas may have pre-existed prior to the shell's expansion.
The observed distributions of cold H{\sc i} gas represent a current snapshot. 
Consequently, quantifying the amount of cold gas formed in situ as a result of stellar-feedback-driven shells remains challenging, even under the relatively simple conditions of the Magellanic Bridge: absence of galactic shearing, little overlap of star-forming regions, and no spiral shocks. 
%


\subsection{Large-scale environment and shell evolution}
The large-scale structure of the Magellanic Bridge is primarily set by the tidal interaction between the LMC and SMC (\citealt{2007MNRAS.381L..11M}), but appears further influenced by local star formation. As we have seen in Figure~\ref{fig:FFT}, the morphology of this portion of the Bridge, particularly in the colder gas, is dominated by large loop structures. These structures are morphologically similar to large supershells observed in many galactic systems, which are usually assumed to be formed by stellar feedback even though they rarely show direct evidence of star formation tracers (\citealt{1997MNRAS.289..225S, 1999MNRAS.302..417S, 1999AJ....118.2797K}).  The presence of colder gas in these filamentary loop structures further indicates a compressive origin and could create an ideal environment for gas condensation in Shell-91.
In Figure~\ref{fig:FFT}, we identify a large-scale structure, MB-Loop~I, with Shell-91 appearing physically connected in projection and kinematics. This suggests a possible physical association and interaction between these structures. 
If the large-scale MB-Loop~I pre-existed, and the local star formation created the shell structure within it, this dense arc within MB-Loop~I may have acted as a physical boundary, causing Shell-91 to preferentially expand into more diffuse regions. 
This interaction likely drove the asymmetric expansion observed in the G$_1$ and G$_4$ components, which are oriented roughly perpendicular to the arc along the loop structure (see Figure~\ref{fig:three_plots}).
%

%
Previous studies have suggested that tidal dwarf galaxies (TDGs) may be forming in the Bridge (\citealt{2015MNRAS.453.3190B, 2020AJ....159...82B}). We find that the large-scale MB-Loop~I and smaller Shell-91 structures are co-located with a TDG candidate (D1 region defined by \citealt{2020AJ....159...82B}), providing a context for re-evaluating TDG formation in the Bridge.
This spatial coincidence is similar to the D3 region, another TDG candidate in the Bridge defined by \citet{2020AJ....159...82B}, located closer to the LMC. In D3, older star clusters are concentrated centrally while younger clusters are distributed along the surrounding shell walls. These young star clusters are associated with expanding H{\sc i} shell structures, indicating that stellar feedback swept up gas and triggered subsequent star formation along the shell boundaries where the gas column density has been enhanced (\citealt{2017MNRAS.472.2975M}).
By analogy, the spatial coincidence of Shell-91 and MB-Loop~I with D1 TDG candidate suggests that these regions may be influenced by expanding shell structures and local stellar feedback, rather than unambiguously tracing tidal dwarf galaxies.

\section{Conclusions and Summary}\label{summary}
In this paper, we present high-resolution (30\arcsec, corresponding to an 8\,pc scale) GASKAP-H{\sc i} observations of H{\sc i} Shell-91 (diameter $\sim$130\,pc) associated with H$\alpha$ shell DEM171 in the Magellanic Bridge. 
The Magellanic Bridge is a favourable environment to investigate the role of stellar feedback on cold H{\sc i} gas formation, as the Bridge is an overall low-density, H{\sc i}-dominated example of the ISM with minimal differential rotation or galactic shearing and fewer overlapping star-forming regions.

To investigate the H{\sc i} multi-phase structure and kinematics of Shell-91, we performed a Gaussian decomposition of the H{\sc i} emission data cube using {\tt ROHSA}. 
We fit the H{\sc i} line profiles with multiple Gaussian components while considering spatial coherence with nearby pixels. 
The decomposed Gaussian G$_2$ component exhibits the narrowest velocity dispersions ($\langle \sigma \rangle < 2.9\,\kms$). Within this narrowest component, we classify the cold gas components with velocity dispersions below 2.5\,\kms. The column density map of the cold H{\sc i} gas shows small-scale structures, and they are located at the shell walls.
The FIR emission traced by \textit{HERschel} 250\,$\mu$m also shows that dust emission is enhanced at the shell boundaries.
While the central cavity of the Shell-91 is ionized, the outskirts of the bubble contain a mixture of cold H{\sc i}, molecular gas traced by CO, and dust.
This shows that the stellar feedback has either promoted the cold gas formation or swept up the cold gas material to the shell boundaries.

However, we also find two additional cold gas clumps outside the shell (one to the north and one to the south), which appear to be part of a larger-scale loop structure. 
To investigate the large-scale view of the cold H{\sc i} gas contents in the Bridge, we applied a Fourier Transform method (\citealt{2024ApJ...961..161M}) to the H{\sc i} emission spectra to map the lower limit of the cold gas column density across the Magellanic Bridge.
This revealed cold H{\sc i} gas structures, including three large-scale loops, MB-Loop~I, II, and III, and shell structures. 
MB-Loop~I and the smaller shell, Shell-91, appear co-located in projection, and the cold gas clumps found outside of Shell-91 seem to correspond to MB-Loop~I structure.

As Shell-91 is connected with the extended arc structure along the MB-Loop~I, this suggests that cold gas may have pre-existed prior to the shell's expansion.
Furthermore, the asymmetric spatial distribution seen in the column density maps of the decomposed Gaussian components, G$_1$ and G$_4$ -- which drive the approaching and receding parts of the shell -- appears to be roughly perpendicular to the arc along the MB-Loop~I. This suggests that the shell's expansion may be influenced by the orientation and presence of large-scale MB-Loop~I structure.
Our observations provide only a single snapshot of the expanding shell; thus it is difficult to disentangle the relative contributions of cold gas formed through shell expansion from that of pre-existing cold gas. 
Quantifying the in situ formation of cold gas within shells remains challenging, even in the relatively simple environment of the Magellanic Bridge, compared to the Milky Way. 
Future studies analyzing a larger sample of shells would be valuable to statistically constrain in situ cold gas formation as a result of stellar feedback.
Additionally, future complementary observations of [C{\sc ii}] or [C{\sc i}] lines will be useful datasets for tracing the cold atomic or ``CO-dark'' molecular phase, where molecular clouds are undetectable in CO emission in low-metallicity environments, such as the Magellanic Bridge (\citealt{2017ApJ...839..107P, 2020A&A...643A.141M}).
Overall, Shell-91 illustrates how stellar feedback operating within a tidally structured medium can shape the large-scale morphology of the gas while promoting the formation of a cold, multiphase ISM.
Also, we show that cold H{\sc i} gas exists over a wide range of scales in the Magellanic Bridge from large-scale loops to small-scale shells (D$\sim$130\,pc). 
We suggest that these cold H{\sc i} maps can also be used as promising targets for future targeted CO observations.

\section*{Acknowledgements}
We thank the anonymous referee for constructive comments and suggestions that helped to improve the manuscript.
This research was funded by the Australian Government through an Australian Research Council Australian Laureate Fellowship (project number FL210100039 awarded to NM-G). B.L. was supported by the National Research Foundation of Korea (NRF) grant funded by the Korea government (MSIT; RS-2026-25498235). \\
This scientific work uses data obtained from Inyarrimanha Ilgari Bundara / the Murchison Radio-astronomy Observatory. We acknowledge the Wajarri Yamaji People as the Traditional Owners and native title holders of the Observatory site. CSIRO’s ASKAP radio telescope is part of the Australia Telescope National Facility (https://ror.org/05qajvd42). Operation of ASKAP is funded by the Australian Government with support from the National Collaborative Research Infrastructure Strategy. ASKAP uses the resources of the Pawsey Supercomputing Research Centre. Establishment of ASKAP, Inyarrimanha Ilgari Bundara, the CSIRO Murchison Radio-astronomy Observatory and the Pawsey Supercomputing Research Centre are initiatives of the Australian Government, with support from the Government of Western Australia and the Science and Industry Endowment Fund. This paper includes archived data obtained through the CSIRO ASKAP Science Data Archive, CASDA (http://data.csiro.au). \\

\emph{Software}: NumPy (\citealt{5725236}), Matplotlib (\citealt{4160265}), Astropy (\citealt{2013A&A...558A..33A}), Scipy (\citealt{2020NatMe..17..261V}), and Photutils (\citealt{larry_bradley_2024_12585239}).

\section*{Data Availability}
The GASKAP-H{\sc i} pilot observations data cube of the Magellanic Bridge is available from the corresponding author upon request.


\bibliographystyle{mnras}
\bibliography{example} 

@ARTICLE{Auclair:2024,
       author = {{Auclair}, Constant and {Allys}, Erwan and {Boulanger}, Fran{\c{c}}ois and {B{\'e}thermin}, Matthieu and {Gkogkou}, Athanasia and {Lagache}, Guilaine and {Marchal}, Antoine and {Miville-Desch{\^e}nes}, Marc-Antoine and {R{\'e}galdo-Saint Blancard}, Bruno and {Richard}, Pablo},
        title = "{Separation of dust emission from the cosmic infrared background in Herschel observations with wavelet phase harmonics}",
      journal = {\aap},
         year = 2024,
        month = jan,
       volume = {681},
          eid = {A1},
        pages = {A1},
          doi = {10.1051/0004-6361/202346814},
archivePrefix = {arXiv},
       eprint = {2305.14419},
 primaryClass = {astro-ph.GA},
       adsurl = {https://ui.adsabs.harvard.edu/abs/2024A&A...681A...1A}
}

@ARTICLE{Sternberg:2014,
       author = {{Sternberg}, Amiel and {Le Petit}, Franck and {Roueff}, Evelyne and {Le Bourlot}, Jacques},
        title = "{H I-to-H$_{2}$ Transitions and H I Column Densities in Galaxy Star-forming Regions}",
      journal = {\apj},
         year = 2014,
        month = jul,
       volume = {790},
       number = {1},
          eid = {10},
        pages = {10},
          doi = {10.1088/0004-637X/790/1/10},
archivePrefix = {arXiv},
       eprint = {1404.5042},
 primaryClass = {astro-ph.GA},
       adsurl = {https://ui.adsabs.harvard.edu/abs/2014ApJ...790...10S}
}

@ARTICLE{Wolfire:2010,
       author = {{Wolfire}, Mark G. and {Hollenbach}, David and {McKee}, Christopher F.},
        title = "{The Dark Molecular Gas}",
      journal = {\apj},
         year = 2010,
        month = jun,
       volume = {716},
       number = {2},
        pages = {1191-1207},
          doi = {10.1088/0004-637X/716/2/1191},
archivePrefix = {arXiv},
       eprint = {1004.5401},
 primaryClass = {astro-ph.GA},
       adsurl = {https://ui.adsabs.harvard.edu/abs/2010ApJ...716.1191W}
}

@ARTICLE{PlanckCollaboration:2011,
       author = {{Planck Collaboration} and {Ade}, P.~A.~R. and {Aghanim}, N. and {Arnaud}, M. and {Ashdown}, M. and {Aumont}, J. and {Baccigalupi}, C. and {Balbi}, A. and {Banday}, A.~J. and {Barreiro}, R.~B. and {Bartlett}, J.~G. and {Battaner}, E. and {Benabed}, K. and {Beno{\^\i}t}, A. and {Bernard}, J.-P. and {Bersanelli}, M. and {Bhatia}, R. and {Bock}, J.~J. and {Bonaldi}, A. and {Bond}, J.~R. and {Borrill}, J. and {Bouchet}, F.~R. and {Boulanger}, F. and {Bucher}, M. and {Burigana}, C. and {Cabella}, P. and {Cardoso}, J.-F. and {Catalano}, A. and {Cay{\'o}n}, L. and {Challinor}, A. and {Chamballu}, A. and {Chiang}, L.-Y. and {Chiang}, C. and {Christensen}, P.~R. and {Clements}, D.~L. and {Colombi}, S. and {Couchot}, F. and {Coulais}, A. and {Crill}, B.~P. and {Cuttaia}, F. and {Dame}, T.~M. and {Danese}, L. and {Davies}, R.~D. and {Davis}, R.~J. and {de Bernardis}, P. and {de Gasperis}, G. and {de Rosa}, A. and {de Zotti}, G. and {Delabrouille}, J. and {Delouis}, J.-M. and {D{\'e}sert}, F.-X. and {Dickinson}, C. and {Dobashi}, K. and {Donzelli}, S. and {Dor{\'e}}, O. and {D{\"o}rl}, U. and {Douspis}, M. and {Dupac}, X. and {Efstathiou}, G. and {En{\ss}lin}, T.~A. and {Eriksen}, H.~K. and {Falgarone}, E. and {Finelli}, F. and {Forni}, O. and {Fosalba}, P. and {Frailis}, M. and {Franceschi}, E. and {Fukui}, Y. and {Galeotta}, S. and {Ganga}, K. and {Giard}, M. and {Giardino}, G. and {Giraud-H{\'e}raud}, Y. and {Gonz{\'a}lez-Nuevo}, J. and {G{\'o}rski}, K.~M. and {Gratton}, S. and {Gregorio}, A. and {Grenier}, I.~A. and {Gruppuso}, A. and {Hansen}, F.~K. and {Harrison}, D. and {Helou}, G. and {Henrot-Versill{\'e}}, S. and {Herranz}, D. and {Hildebrandt}, S.~R. and {Hivon}, E. and {Hobson}, M. and {Holmes}, W.~A. and {Hovest}, W. and {Hoyland}, R.~J. and {Huffenberger}, K.~M. and {Jaffe}, A.~H. and {Jones}, W.~C. and {Juvela}, M. and {Kawamura}, A. and {Keih{\"a}nen}, E. and {Keskitalo}, R. and {Kisner}, T.~S. and {Kneissl}, R. and {Knox}, L. and {Kurki-Suonio}, H. and {Lagache}, G. and {Lamarre}, J.-M. and {Lasenby}, A. and {Laureijs}, R.~J. and {Lawrence}, C.~R. and {Leach}, S. and {Leonardi}, R. and {Leroy}, C. and {Lilje}, P.~B. and {Linden-V{\o}rnle}, M. and {L{\'o}pez-Caniego}, M. and {Lubin}, P.~M. and {Mac{\'\i}as-P{\'e}rez}, J.~F. and {MacTavish}, C.~J. and {Maffei}, B. and {Maino}, D. and {Mandolesi}, N. and {Mann}, R. and {Maris}, M. and {Martin}, P. and {Mart{\'\i}nez-Gonz{\'a}lez}, E. and {Masi}, S. and {Matarrese}, S. and {Matthai}, F. and {Mazzotta}, P. and {McGehee}, P. and {Meinhold}, P.~R. and {Melchiorri}, A. and {Mendes}, L. and {Mennella}, A. and {Miville-Desch{\^e}nes}, M.-A. and {Moneti}, A. and {Montier}, L. and {Morgante}, G. and {Mortlock}, D. and {Munshi}, D. and {Murphy}, A. and {Naselsky}, P. and {Natoli}, P. and {Netterfield}, C.~B. and {N{\o}rgaard-Nielsen}, H.~U. and {Noviello}, F. and {Novikov}, D. and {Novikov}, I. and {O'Dwyer}, I.~J. and {Onishi}, T. and {Osborne}, S. and {Pajot}, F. and {Paladini}, R. and {Paradis}, D. and {Pasian}, F. and {Patanchon}, G. and {Perdereau}, O. and {Perotto}, L. and {Perrotta}, F. and {Piacentini}, F. and {Piat}, M. and {Plaszczynski}, S. and {Pointecouteau}, E. and {Polenta}, G. and {Ponthieu}, N. and {Poutanen}, T. and {Pr{\'e}zeau}, G. and {Prunet}, S. and {Puget}, J.-L. and {Reach}, W.~T. and {Reinecke}, M. and {Renault}, C. and {Ricciardi}, S. and {Riller}, T. and {Ristorcelli}, I. and {Rocha}, G. and {Rosset}, C. and {Rowan-Robinson}, M. and {Rubi{\~n}o-Mart{\'\i}n}, J.~A. and {Rusholme}, B. and {Sandri}, M. and {Santos}, D. and {Savini}, G. and {Scott}, D. and {Seiffert}, M.~D. and {Shellard}, P. and {Smoot}, G.~F. and {Starck}, J.-L. and {Stivoli}, F. and {Stolyarov}, V. and {Stompor}, R. and {Sudiwala}, R. and {Sygnet}, J.-F. and {Tauber}, J.~A. and {Terenzi}, L. and {Toffolatti}, L. and {Tomasi}, M. and {Torre}, J.-P. and {Tristram}, M. and {Tuovinen}, J. and {Umana}, G. and {Valenziano}, L. and {Vielva}, P.},
        title = "{Planck early results. XIX. All-sky temperature and dust optical depth from Planck and IRAS. Constraints on the ``dark gas'' in our Galaxy}",
      journal = {\aap},
         year = 2011,
        month = dec,
       volume = {536},
          eid = {A19},
        pages = {A19},
          doi = {10.1051/0004-6361/201116479},
archivePrefix = {arXiv},
       eprint = {1101.2029},
 primaryClass = {astro-ph.GA},
       adsurl = {https://ui.adsabs.harvard.edu/abs/2011A&A...536A..19P}
}

@ARTICLE{Boulanger:1996,
       author = {{Boulanger}, F. and {Abergel}, A. and {Bernard}, J.-P. and {Burton}, W.~B. and {Desert}, F.-X. and {Hartmann}, D. and {Lagache}, G. and {Puget}, J.-L.},
        title = "{The dust/gas correlation at high Galactic latitude.}",
      journal = {\aap},
         year = 1996,
        month = aug,
       volume = {312},
        pages = {256-262},
       adsurl = {https://ui.adsabs.harvard.edu/abs/1996A&A...312..256B}
}

@ARTICLE{Scalo:2004,
       author = {{Scalo}, John and {Elmegreen}, Bruce G.},
        title = "{Interstellar Turbulence II: Implications and Effects}",
      journal = {\araa},
         year = 2004,
        month = sep,
       volume = {42},
       number = {1},
        pages = {275-316},
          doi = {10.1146/annurev.astro.42.120403.143327},
archivePrefix = {arXiv},
       eprint = {astro-ph/0404452},
 primaryClass = {astro-ph},
       adsurl = {https://ui.adsabs.harvard.edu/abs/2004ARA&A..42..275S}
}

@ARTICLE{Elmegreen:2004,
       author = {{Elmegreen}, Bruce G. and {Scalo}, John},
        title = "{Interstellar Turbulence I: Observations and Processes}",
      journal = {\araa},
         year = 2004,
        month = sep,
       volume = {42},
       number = {1},
        pages = {211-273},
          doi = {10.1146/annurev.astro.41.011802.094859},
archivePrefix = {arXiv},
       eprint = {astro-ph/0404451},
 primaryClass = {astro-ph},
       adsurl = {https://ui.adsabs.harvard.edu/abs/2004ARA&A..42..211E}
}

@ARTICLE{2014A&A...567A..16S,
       author = {{Saury}, E. and {Miville-Desch{\^e}nes}, M. -A. and {Hennebelle}, P. and {Audit}, E. and {Schmidt}, W.},
        title = "{The structure of the thermally bistable and turbulent atomic gas in the local interstellar medium}",
      journal = {\aap},
         year = 2014,
        month = jul,
       volume = {567},
          eid = {A16},
        pages = {A16},
          doi = {10.1051/0004-6361/201321113},
archivePrefix = {arXiv},
       eprint = {1301.3446},
 primaryClass = {astro-ph.GA},
       adsurl = {https://ui.adsabs.harvard.edu/abs/2014A&A...567A..16S}
}

@ARTICLE{2021A&A...646A..16R,
       author = {{Ramachandran}, V. and {Oskinova}, L.~M. and {Hamann}, W. -R.},
        title = "{Discovery of O stars in the tidal Magellanic Bridge. Stellar parameters, abundances, and feedback of the nearest metal-poor massive stars and their implication for the Magellanic System ecology}",
      journal = {\aap},
         year = 2021,
        month = feb,
       volume = {646},
          eid = {A16},
        pages = {A16},
          doi = {10.1051/0004-6361/202039486},
archivePrefix = {arXiv},
       eprint = {2011.08006},
 primaryClass = {astro-ph.GA},
       adsurl = {https://ui.adsabs.harvard.edu/abs/2021A&A...646A..16R}
}

@ARTICLE{2014ApJ...785..162C,
       author = {{Chen}, C. -H. Rosie and {Indebetouw}, Remy and {Muller}, Erik and {Kawamura}, Akiko and {Gordon}, Karl D. and {Sewi{\l}o}, Marta and {Whitney}, Barbara A. and {Fukui}, Yasuo and {Madden}, Suzanne C. and {Meade}, Marilyn R. and {Meixner}, Margaret and {Oliveira}, Joana M. and {Robitaille}, Thomas P. and {Seale}, Jonathan P. and {Shiao}, Bernie and {van Loon}, Jacco Th.},
        title = "{Spitzer View of Massive Star Formation in the Tidally Stripped Magellanic Bridge}",
      journal = {\apj},
         year = 2014,
        month = apr,
       volume = {785},
       number = {2},
          eid = {162},
        pages = {162},
          doi = {10.1088/0004-637X/785/2/162},
archivePrefix = {arXiv},
       eprint = {1403.0618},
 primaryClass = {astro-ph.GA},
       adsurl = {https://ui.adsabs.harvard.edu/abs/2014ApJ...785..162C}
}

@ARTICLE{marchal_2019,
       author = {{Marchal}, Antoine and {Miville-Desch{\^e}nes}, Marc-Antoine and {Orieux}, Fran{\c{c}}ois and {Gac}, Nicolas and {Soussen}, Charles and {Lesot}, Marie-Jeanne and {d'Allonnes}, Adrien Revault and {Salom{\'e}}, Quentin},
        title = "{ROHSA: Regularized Optimization for Hyper-Spectral Analysis. Application to phase separation of 21 cm data}",
      journal = {\aap},
         year = 2019,
        month = jun,
       volume = {626},
          eid = {A101},
        pages = {A101},
          doi = {10.1051/0004-6361/201935335},
archivePrefix = {arXiv},
       eprint = {1905.00658},
 primaryClass = {astro-ph.GA},
       adsurl = {https://ui.adsabs.harvard.edu/abs/2019A&A...626A.101M}
}

@ARTICLE{2017MNRAS.472.2975M,
       author = {{Mackey}, A.~D. and {Koposov}, S.~E. and {Da Costa}, G.~S. and {Belokurov}, V. and {Erkal}, D. and {Fraternali}, F. and {McClure-Griffiths}, N.~M. and {Fraser}, M.},
        title = "{Structured star formation in the Magellanic inter-Cloud region}",
      journal = {\mnras},
         year = 2017,
        month = dec,
       volume = {472},
       number = {3},
        pages = {2975-2989},
          doi = {10.1093/mnras/stx2035},
archivePrefix = {arXiv},
       eprint = {1708.04363},
 primaryClass = {astro-ph.GA},
       adsurl = {https://ui.adsabs.harvard.edu/abs/2017MNRAS.472.2975M}
}

@ARTICLE{2007PASA...24...69M,
       author = {{Muller}, E. and {Parker}, Q.~A.},
        title = "{H{\ensuremath{\alpha}} Emission from the Magellanic Bridge}",
      journal = {\pasa},
         year = 2007,
        month = jul,
       volume = {24},
       number = {2},
        pages = {69-76},
          doi = {10.1071/AS07010},
archivePrefix = {arXiv},
       eprint = {0706.4037},
 primaryClass = {astro-ph},
       adsurl = {https://ui.adsabs.harvard.edu/abs/2007PASA...24...69M}
}

@ARTICLE{2003MNRAS.338..609M,
       author = {{Muller}, E. and {Staveley-Smith}, L. and {Zealey}, W.~J.},
        title = "{Detection of carbon monoxide within the Magellanic Bridge}",
      journal = {\mnras},
         year = 2003,
        month = jan,
       volume = {338},
       number = {3},
        pages = {609-615},
          doi = {10.1046/j.1365-8711.2003.06062.x},
archivePrefix = {arXiv},
       eprint = {astro-ph/0209523},
 primaryClass = {astro-ph},
       adsurl = {https://ui.adsabs.harvard.edu/abs/2003MNRAS.338..609M}
}

@ARTICLE{2020A&A...641A..97V,
       author = {{Valdivia-Mena}, M.~T. and {Rubio}, M. and {Bolatto}, A.~D. and {Salda{\~n}o}, H.~P. and {Verdugo}, C.},
        title = "{ALMA resolves molecular clouds in metal-poor Magellanic Bridge A}",
      journal = {\aap},
         year = 2020,
        month = sep,
       volume = {641},
          eid = {A97},
        pages = {A97},
          doi = {10.1051/0004-6361/201937232},
archivePrefix = {arXiv},
       eprint = {2007.01319},
 primaryClass = {astro-ph.GA},
       adsurl = {https://ui.adsabs.harvard.edu/abs/2020A&A...641A..97V}
}

@ARTICLE{2006ApJ...643L.107M,
       author = {{Mizuno}, N. and {Muller}, E. and {Maeda}, H. and {Kawamura}, A. and {Minamidani}, T. and {Onishi}, T. and {Mizuno}, A. and {Fukui}, Y.},
        title = "{Detection of Molecular Clouds in the Magellanic Bridge: Candidate Star Formation Sites in a Nearby Low-Metallicity System}",
      journal = {\apjl},
         year = 2006,
        month = jun,
       volume = {643},
       number = {2},
        pages = {L107-L110},
          doi = {10.1086/505298},
       adsurl = {https://ui.adsabs.harvard.edu/abs/2006ApJ...643L.107M}
}

@ARTICLE{2011ApJ...741...85D,
       author = {{Dawson}, J.~R. and {McClure-Griffiths}, N.~M. and {Dickey}, John M. and {Fukui}, Y.},
        title = "{Molecular Clouds in Supershells: A Case Study of Three Objects in the Walls of GSH 287+04-17 and GSH 277+00+36}",
      journal = {\apj},
         year = 2011,
        month = nov,
       volume = {741},
       number = {2},
          eid = {85},
        pages = {85},
          doi = {10.1088/0004-637X/741/2/85},
archivePrefix = {arXiv},
       eprint = {1108.3882},
 primaryClass = {astro-ph.GA},
       adsurl = {https://ui.adsabs.harvard.edu/abs/2011ApJ...741...85D}
}

@ARTICLE{2003MNRAS.339..105M,
       author = {{Muller}, E. and {Staveley-Smith}, L. and {Zealey}, W. and {Stanimirovi{\'c}}, S.},
        title = "{High-resolution HI observations of the Western Magellanic Bridge}",
      journal = {\mnras},
         year = 2003,
        month = feb,
       volume = {339},
       number = {1},
        pages = {105-124},
          doi = {10.1046/j.1365-8711.2003.06147.x},
archivePrefix = {arXiv},
       eprint = {astro-ph/0210615},
 primaryClass = {astro-ph},
       adsurl = {https://ui.adsabs.harvard.edu/abs/2003MNRAS.339..105M}
}

@ARTICLE{2013ApJ...763...56D,
       author = {{Dawson}, J.~R. and {McClure-Griffiths}, N.~M. and {Wong}, T. and {Dickey}, John M. and {Hughes}, A. and {Fukui}, Y. and {Kawamura}, A.},
        title = "{Supergiant Shells and Molecular Cloud Formation in the Large Magellanic Cloud}",
      journal = {\apj},
         year = 2013,
        month = jan,
       volume = {763},
       number = {1},
          eid = {56},
        pages = {56},
          doi = {10.1088/0004-637X/763/1/56},
archivePrefix = {arXiv},
       eprint = {1211.7119},
 primaryClass = {astro-ph.GA},
       adsurl = {https://ui.adsabs.harvard.edu/abs/2013ApJ...763...56D}
}

@ARTICLE{2000A&A...354..787M,
       author = {{Marx-Zimmer}, M. and {Herbstmeier}, U. and {Dickey}, J.~M. and {Zimmer}, F. and {Staveley-Smith}, L. and {Mebold}, U.},
        title = "{A study of the cool gas in the Large Magellanic Cloud. I. Properties of the cool atomic phase - a third H i absorption survey}",
      journal = {\aap},
         year = 2000,
        month = feb,
       volume = {354},
        pages = {787-801},
       adsurl = {https://ui.adsabs.harvard.edu/abs/2000A&A...354..787M}
}

@ARTICLE{2009ApJ...690L..76G,
       author = {{Gordon}, K.~D. and {Bot}, C. and {Muller}, E. and {Misselt}, K.~A. and {Bolatto}, A. and {Bernard}, J. -P. and {Reach}, W. and {Engelbracht}, C.~W. and {Babler}, B. and {Bracker}, S. and {Block}, M. and {Clayton}, G.~C. and {Hora}, J. and {Indebetouw}, R. and {Israel}, F.~P. and {Li}, A. and {Madden}, S. and {Meade}, M. and {Meixner}, M. and {Sewilo}, M. and {Shiao}, B. and {Smith}, L.~J. and {van Loon}, J. Th. and {Whitney}, B.~A.},
        title = "{The Dust-to-Gas Ratio in the Small Magellanic Cloud Tail}",
      journal = {\apjl},
         year = 2009,
        month = jan,
       volume = {690},
       number = {1},
        pages = {L76-L80},
          doi = {10.1088/0004-637X/690/1/L76},
archivePrefix = {arXiv},
       eprint = {0811.2789},
 primaryClass = {astro-ph},
       adsurl = {https://ui.adsabs.harvard.edu/abs/2009ApJ...690L..76G}
}

@ARTICLE{1986MNRAS.223..317M,
       author = {{Meaburn}, J.},
        title = "{Young-star formation in the Magellanic H I bridge.}",
      journal = {\mnras},
         year = 1986,
        month = nov,
       volume = {223},
        pages = {317-321},
          doi = {10.1093/mnras/223.2.317},
       adsurl = {https://ui.adsabs.harvard.edu/abs/1986MNRAS.223..317M}
}

@ARTICLE{2002ApJ...578..176M,
       author = {{McClure-Griffiths}, N.~M. and {Dickey}, John M. and {Gaensler}, B.~M. and {Green}, A.~J.},
        title = "{The Galactic Distribution of Large H I Shells}",
      journal = {\apj},
         year = 2002,
        month = oct,
       volume = {578},
       number = {1},
        pages = {176-193},
          doi = {10.1086/342470},
archivePrefix = {arXiv},
       eprint = {astro-ph/0206358},
 primaryClass = {astro-ph},
       adsurl = {https://ui.adsabs.harvard.edu/abs/2002ApJ...578..176M}
}

@ARTICLE{2003ApJ...594..833M,
       author = {{McClure-Griffiths}, N.~M. and {Dickey}, John M. and {Gaensler}, B.~M. and {Green}, A.~J.},
        title = "{Loops, Drips, and Walls in the Galactic Chimney GSH 277+00+36}",
      journal = {\apj},
         year = 2003,
        month = sep,
       volume = {594},
       number = {2},
        pages = {833-843},
          doi = {10.1086/377152},
       adsurl = {https://ui.adsabs.harvard.edu/abs/2003ApJ...594..833M}
}

@ARTICLE{1999AJ....118.2797K,
       author = {{Kim}, Sungeun and {Dopita}, Michael A. and {Staveley-Smith}, Lister and {Bessell}, Michael S.},
        title = "{H I Shells in the Large Magellanic Cloud}",
      journal = {\aj},
         year = 1999,
        month = dec,
       volume = {118},
       number = {6},
        pages = {2797-2823},
          doi = {10.1086/301116},
       adsurl = {https://ui.adsabs.harvard.edu/abs/1999AJ....118.2797K}
}

@ARTICLE{2020AJ....160...66P,
       author = {{Pokhrel}, Nau Raj and {Simpson}, Caroline E. and {Bagetakos}, Ioannis},
        title = "{A Catalog of Holes and Shells in the Interstellar Medium of the LITTLE THINGS Dwarf Galaxies}",
      journal = {\aj},
         year = 2020,
        month = aug,
       volume = {160},
       number = {2},
          eid = {66},
        pages = {66},
          doi = {10.3847/1538-3881/ab9bfa},
archivePrefix = {arXiv},
       eprint = {2006.01735},
 primaryClass = {astro-ph.GA},
       adsurl = {https://ui.adsabs.harvard.edu/abs/2020AJ....160...66P}
}

@ARTICLE{2005MNRAS.362..689P,
       author = {{Parker}, Quentin A. and {Phillipps}, S. and {Pierce}, M.~J. and {Hartley}, M. and {Hambly}, N.~C. and {Read}, M.~A. and {MacGillivray}, H.~T. and {Tritton}, S.~B. and {Cass}, C.~P. and {Cannon}, R.~D. and {Cohen}, M. and {Drew}, J.~E. and {Frew}, D.~J. and {Hopewell}, E. and {Mader}, S. and {Malin}, D.~F. and {Masheder}, M.~R.~W. and {Morgan}, D.~H. and {Morris}, R.~A.~H. and {Russeil}, D. and {Russell}, K.~S. and {Walker}, R.~N.~F.},
        title = "{The AAO/UKST SuperCOSMOS H{\ensuremath{\alpha}} survey}",
      journal = {\mnras},
         year = 2005,
        month = sep,
       volume = {362},
       number = {2},
        pages = {689-710},
          doi = {10.1111/j.1365-2966.2005.09350.x},
archivePrefix = {arXiv},
       eprint = {astro-ph/0506599},
 primaryClass = {astro-ph},
       adsurl = {https://ui.adsabs.harvard.edu/abs/2005MNRAS.362..689P}
}

@ARTICLE{2024ApJ...961..161M,
       author = {{Marchal}, Antoine and {Martin}, Peter G. and {Miville-Desch{\^e}nes}, Marc-Antoine and {McClure-Griffiths}, Naomi M. and {Lynn}, Callum and {Bracco}, Andrea and {Vujeva}, Luka},
        title = "{Mapping a Lower Limit on the Mass Fraction of the Cold Neutral Medium Using Fourier-transformed H I 21 cm Emission Line Spectra: Application to the DRAO Deep Field from DHIGLS and the HI4PI Survey}",
      journal = {\apj},
         year = 2024,
        month = feb,
       volume = {961},
       number = {2},
          eid = {161},
        pages = {161},
          doi = {10.3847/1538-4357/ad0f21},
archivePrefix = {arXiv},
       eprint = {2311.15122},
 primaryClass = {astro-ph.GA},
       adsurl = {https://ui.adsabs.harvard.edu/abs/2024ApJ...961..161M}
}

@ARTICLE{2020AJ....159...82B,
       author = {{Bica}, Eduardo and {Westera}, Pieter and {Kerber}, Leandro de O. and {Dias}, Bruno and {Maia}, Francisco and {Santos}, Jo{\~a}o F.~C., Jr. and {Barbuy}, Beatriz and {Oliveira}, Raphael A.~P.},
        title = "{An Updated Small Magellanic Cloud and Magellanic Bridge Catalog of Star Clusters, Associations, and Related Objects}",
      journal = {\aj},
         year = 2020,
        month = mar,
       volume = {159},
       number = {3},
          eid = {82},
        pages = {82},
          doi = {10.3847/1538-3881/ab6595},
archivePrefix = {arXiv},
       eprint = {1907.08642},
 primaryClass = {astro-ph.GA},
       adsurl = {https://ui.adsabs.harvard.edu/abs/2020AJ....159...82B}
}

@ARTICLE{2009ApJS..181..398M,
       author = {{McClure-Griffiths}, N.~M. and {Pisano}, D.~J. and {Calabretta}, M.~R. and {Ford}, H. Alyson and {Lockman}, Felix J. and {Staveley-Smith}, L. and {Kalberla}, P.~M.~W. and {Bailin}, J. and {Dedes}, L. and {Janowiecki}, S. and {Gibson}, B.~K. and {Murphy}, T. and {Nakanishi}, H. and {Newton-McGee}, K.},
        title = "{Gass: The Parkes Galactic All-Sky Survey. I. Survey Description, Goals, and Initial Data Release}",
      journal = {\apjs},
         year = 2009,
        month = apr,
       volume = {181},
       number = {2},
        pages = {398-412},
          doi = {10.1088/0067-0049/181/2/398},
archivePrefix = {arXiv},
       eprint = {0901.1159},
 primaryClass = {astro-ph.GA},
       adsurl = {https://ui.adsabs.harvard.edu/abs/2009ApJS..181..398M}
}

@ARTICLE{2021PASA...38....9H,
       author = {{Hotan}, A.~W. and {Bunton}, J.~D. and {Chippendale}, A.~P. and {Whiting}, M. and {Tuthill}, J. and {Moss}, V.~A. and {McConnell}, D. and {Amy}, S.~W. and {Huynh}, M.~T. and {Allison}, J.~R. and {Anderson}, C.~S. and {Bannister}, K.~W. and {Bastholm}, E. and {Beresford}, R. and {Bock}, D.~C. -J. and {Bolton}, R. and {Chapman}, J.~M. and {Chow}, K. and {Collier}, J.~D. and {Cooray}, F.~R. and {Cornwell}, T.~J. and {Diamond}, P.~J. and {Edwards}, P.~G. and {Feain}, I.~J. and {Franzen}, T.~M.~O. and {George}, D. and {Gupta}, N. and {Hampson}, G.~A. and {Harvey-Smith}, L. and {Hayman}, D.~B. and {Heywood}, I. and {Jacka}, C. and {Jackson}, C.~A. and {Jackson}, S. and {Jeganathan}, K. and {Johnston}, S. and {Kesteven}, M. and {Kleiner}, D. and {Koribalski}, B.~S. and {Lee-Waddell}, K. and {Lenc}, E. and {Lensson}, E.~S. and {Mackay}, S. and {Mahony}, E.~K. and {McClure-Griffiths}, N.~M. and {McConigley}, R. and {Mirtschin}, P. and {Ng}, A.~K. and {Norris}, R.~P. and {Pearce}, S.~E. and {Phillips}, C. and {Pilawa}, M.~A. and {Raja}, W. and {Reynolds}, J.~E. and {Roberts}, P. and {Roxby}, D.~N. and {Sadler}, E.~M. and {Shields}, M. and {Schinckel}, A.~E.~T. and {Serra}, P. and {Shaw}, R.~D. and {Sweetnam}, T. and {Troup}, E.~R. and {Tzioumis}, A. and {Voronkov}, M.~A. and {Westmeier}, T.},
        title = "{Australian square kilometre array pathfinder: I. system description}",
      journal = {\pasa},
         year = 2021,
        month = mar,
       volume = {38},
          eid = {e009},
        pages = {e009},
          doi = {10.1017/pasa.2021.1},
archivePrefix = {arXiv},
       eprint = {2102.01870},
 primaryClass = {astro-ph.IM},
       adsurl = {https://ui.adsabs.harvard.edu/abs/2021PASA...38....9H}
}

@ARTICLE{2022PASA...39....5P,
       author = {{Pingel}, N.~M. and {Dempsey}, J. and {McClure-Griffiths}, N.~M. and {Dickey}, J.~M. and {Jameson}, K.~E. and {Arce}, H. and {Anglada}, G. and {Bland-Hawthorn}, J. and {Breen}, S.~L. and {Buckland-Willis}, F. and {Clark}, S.~E. and {Dawson}, J.~R. and {D{\'e}nes}, H. and {Di Teodoro}, E.~M. and {For}, B. -Q. and {Foster}, Tyler J. and {G{\'o}mez}, J.~F. and {Imai}, H. and {Joncas}, G. and {Kim}, C. -G. and {Lee}, M. -Y. and {Lynn}, C. and {Leahy}, D. and {Ma}, Y.~K. and {Marchal}, A. and {McConnell}, D. and {Miville-Desch{\`e}nes}, M. -A. and {Moss}, V.~A. and {Murray}, C.~E. and {Nidever}, D. and {Peek}, J. and {Stanimirovi{\'c}}, S. and {Staveley-Smith}, L. and {Tepper-Garcia}, T. and {Tremblay}, C.~D. and {Uscanga}, L. and {van Loon}, J. Th. and {V{\'a}zquez-Semadeni}, E. and {Allison}, J.~R. and {Anderson}, C.~S. and {Ball}, Lewis and {Bell}, M. and {Bock}, D.~C. -J. and {Bunton}, J. and {Cooray}, F.~R. and {Cornwell}, T. and {Koribalski}, B.~S. and {Gupta}, N. and {Hayman}, D.~B. and {Harvey-Smith}, L. and {Lee-Waddell}, K. and {Ng}, A. and {Phillips}, C.~J. and {Voronkov}, M. and {Westmeier}, T. and {Whiting}, M.~T.},
        title = "{GASKAP-HI pilot survey science I: ASKAP zoom observations of HI emission in the Small Magellanic Cloud}",
      journal = {\pasa},
         year = 2022,
        month = feb,
       volume = {39},
          eid = {e005},
        pages = {e005},
          doi = {10.1017/pasa.2021.59},
archivePrefix = {arXiv},
       eprint = {2111.05339},
 primaryClass = {astro-ph.GA},
       adsurl = {https://ui.adsabs.harvard.edu/abs/2022PASA...39....5P}
}

@ARTICLE{2013PASA...30....3D,
       author = {{Dickey}, John M. and {McClure-Griffiths}, Naomi and {Gibson}, Steven J. and {G{\'o}mez}, Jos{\'e} F. and {Imai}, Hiroshi and {Jones}, Paul and {Stanimirovi{\'c}}, Sne{\v{z}}ana and {Van Loon}, Jacco Th. and {Walsh}, Andrew and {Alberdi}, A. and {Anglada}, G. and {Uscanga}, L. and {Arce}, H. and {Bailey}, M. and {Begum}, A. and {Wakker}, B. and {Bekhti}, N. Ben and {Kalberla}, P. and {Winkel}, B. and {Bekki}, K. and {For}, B. -Q. and {Staveley-Smith}, L. and {Westmeier}, T. and {Burton}, M. and {Cunningham}, M. and {Dawson}, J. and {Ellingsen}, S. and {Diamond}, P. and {Green}, J.~A. and {Hill}, A.~S. and {Koribalski}, B. and {McConnell}, D. and {Rathborne}, J. and {Voronkov}, M. and {Douglas}, K.~A. and {English}, J. and {Ford}, H. Alyson and {Lockman}, F.~J. and {Foster}, T. and {Gomez}, Y. and {Green}, A. and {Bland-Hawthorn}, J. and {Gulyaev}, S. and {Hoare}, M. and {Joncas}, G. and {Kang}, J. -H. and {Kerton}, C.~R. and {Koo}, B. -C. and {Leahy}, D. and {Lo}, N. and {Migenes}, V. and {Nakashima}, J. and {Zhang}, Y. and {Nidever}, D. and {Peek}, J.~E.~G. and {Tafoya}, D. and {Tian}, W. and {Wu}, D.},
        title = "{GASKAP-The Galactic ASKAP Survey}",
      journal = {\pasa},
         year = 2013,
        month = jan,
       volume = {30},
          eid = {e003},
        pages = {e003},
          doi = {10.1017/pasa.2012.003},
archivePrefix = {arXiv},
       eprint = {1207.0891},
 primaryClass = {astro-ph.GA},
       adsurl = {https://ui.adsabs.harvard.edu/abs/2013PASA...30....3D}
}

@ARTICLE{2014ApJ...796..123F,
       author = {{Fujii}, Kosuke and {Minamidani}, Tetsuhiro and {Mizuno}, Norikazu and {Onishi}, Toshikazu and {Kawamura}, Akiko and {Muller}, Erik and {Dawson}, Joanne and {Tatematsu}, Ken'ichi and {Hasegawa}, Tetsuo and {Tosaki}, Tomoka and {Miura}, Rie E. and {Muraoka}, Kazuyuki and {Sakai}, Takeshi and {Tsukagoshi}, Takashi and {Tanaka}, Kunihiko and {Ezawa}, Hajime and {Fukui}, Yasuo},
        title = "{Dense Molecular Clumps Associated with the Large Magellanic Cloud Supergiant Shells LMC 4 and LMC 5}",
      journal = {\apj},
         year = 2014,
        month = dec,
       volume = {796},
       number = {2},
          eid = {123},
        pages = {123},
          doi = {10.1088/0004-637X/796/2/123},
archivePrefix = {arXiv},
       eprint = {1411.0097},
 primaryClass = {astro-ph.GA},
       adsurl = {https://ui.adsabs.harvard.edu/abs/2014ApJ...796..123F}
}

@ARTICLE{2022ApJ...937...81T,
       author = {{Taank}, Mukesh and {Marchal}, Antoine and {Martin}, Peter G. and {Vujeva}, Luka},
        title = "{Mapping the Thermal Condensation of Diffuse H I in the North Celestial Pole Loop}",
      journal = {\apj},
         year = 2022,
        month = oct,
       volume = {937},
       number = {2},
          eid = {81},
        pages = {81},
          doi = {10.3847/1538-4357/ac8b86},
archivePrefix = {arXiv},
       eprint = {2209.14998},
 primaryClass = {astro-ph.GA},
       adsurl = {https://ui.adsabs.harvard.edu/abs/2022ApJ...937...81T}
}

@ARTICLE{2014PASJ...66....4M,
       author = {{Muller}, Erik and {Mizuno}, Norikazu and {Minamidani}, Tetsuhiro and {Kawamura}, Akiko and {Rosie Chen}, C. -H. and {Indebetouw}, Remy and {Enokiya}, Rei and {Fukui}, Yasuo and {Gordon}, Karl and {Hayakawa}, Takahiro and {Mizuno}, Yoji and {Murai}, Miyuki and {Okuda}, Takeshi and {Onishi}, Toshikazu and {Tachihara}, Kengo and {Takekoshi}, Tatsuya and {Yamamoto}, Hiroaki and {Yoshiike}, Satoshi},
        title = "{Unusually bright $^{12}$CO(3-2) condensations in the tidally perturbed Small Magellanic Cloud ``tail''}",
      journal = {\pasj},
         year = 2014,
        month = feb,
       volume = {66},
       number = {1},
          eid = {4},
        pages = {4},
          doi = {10.1093/pasj/pst006},
       adsurl = {https://ui.adsabs.harvard.edu/abs/2014PASJ...66....4M}
}

@ARTICLE{2020ApJ...905...95K,
       author = {{Kobayashi}, Masato I.~N. and {Inoue}, Tsuyoshi and {Inutsuka}, Shu-ichiro and {Tomida}, Kengo and {Iwasaki}, Kazunari and {Tanaka}, Kei E.~I.},
        title = "{Bimodal Behavior and Convergence Requirement in Macroscopic Properties of the Multiphase Interstellar Medium Formed by Atomic Converging Flows}",
      journal = {\apj},
         year = 2020,
        month = dec,
       volume = {905},
       number = {2},
          eid = {95},
        pages = {95},
          doi = {10.3847/1538-4357/abc5be},
archivePrefix = {arXiv},
       eprint = {2010.12368},
 primaryClass = {astro-ph.GA},
       adsurl = {https://ui.adsabs.harvard.edu/abs/2020ApJ...905...95K}
}

@ARTICLE{2013AJ....146...62M,
       author = {{Meixner}, M. and {Panuzzo}, P. and {Roman-Duval}, J. and {Engelbracht}, C. and {Babler}, B. and {Seale}, J. and {Hony}, S. and {Montiel}, E. and {Sauvage}, M. and {Gordon}, K. and {Misselt}, K. and {Okumura}, K. and {Chanial}, P. and {Beck}, T. and {Bernard}, J. -P. and {Bolatto}, A. and {Bot}, C. and {Boyer}, M.~L. and {Carlson}, L.~R. and {Clayton}, G.~C. and {Chen}, C. -H.~R. and {Cormier}, D. and {Fukui}, Y. and {Galametz}, M. and {Galliano}, F. and {Hora}, J.~L. and {Hughes}, A. and {Indebetouw}, R. and {Israel}, F.~P. and {Kawamura}, A. and {Kemper}, F. and {Kim}, S. and {Kwon}, E. and {Lebouteiller}, V. and {Li}, A. and {Long}, K.~S. and {Madden}, S.~C. and {Matsuura}, M. and {Muller}, E. and {Oliveira}, J.~M. and {Onishi}, T. and {Otsuka}, M. and {Paradis}, D. and {Poglitsch}, A. and {Reach}, W.~T. and {Robitaille}, T.~P. and {Rubio}, M. and {Sargent}, B. and {Sewi{\l}o}, M. and {Skibba}, R. and {Smith}, L.~J. and {Srinivasan}, S. and {Tielens}, A.~G.~G.~M. and {van Loon}, J. Th. and {Whitney}, B.},
        title = "{The HERSCHEL Inventory of The Agents of Galaxy Evolution in the Magellanic Clouds, a Herschel Open Time Key Program}",
      journal = {\aj},
         year = 2013,
        month = sep,
       volume = {146},
       number = {3},
          eid = {62},
        pages = {62},
          doi = {10.1088/0004-6256/146/3/62},
       adsurl = {https://ui.adsabs.harvard.edu/abs/2013AJ....146...62M}
}

@ARTICLE{2014ApJ...797...85G,
       author = {{Gordon}, Karl D. and {Roman-Duval}, Julia and {Bot}, Caroline and {Meixner}, Margaret and {Babler}, Brian and {Bernard}, Jean-Philippe and {Bolatto}, Alberto and {Boyer}, Martha L. and {Clayton}, Geoffrey C. and {Engelbracht}, Charles and {Fukui}, Yasuo and {Galametz}, Maud and {Galliano}, Frederic and {Hony}, Sacha and {Hughes}, Annie and {Indebetouw}, Remy and {Israel}, Frank P. and {Jameson}, Katherine and {Kawamura}, Akiko and {Lebouteiller}, Vianney and {Li}, Aigen and {Madden}, Suzanne C. and {Matsuura}, Mikako and {Misselt}, Karl and {Montiel}, Edward and {Okumura}, K. and {Onishi}, Toshikazu and {Panuzzo}, Pasquale and {Paradis}, Deborah and {Rubio}, Monica and {Sandstrom}, Karin and {Sauvage}, Marc and {Seale}, Jonathan and {Sewi{\l}o}, Marta and {Tchernyshyov}, Kirill and {Skibba}, Ramin},
        title = "{Dust and Gas in the Magellanic Clouds from the HERITAGE Herschel Key Project. I. Dust Properties and Insights into the Origin of the Submillimeter Excess Emission}",
      journal = {\apj},
         year = 2014,
        month = dec,
       volume = {797},
       number = {2},
          eid = {85},
        pages = {85},
          doi = {10.1088/0004-637X/797/2/85},
archivePrefix = {arXiv},
       eprint = {1406.6066},
 primaryClass = {astro-ph.GA},
       adsurl = {https://ui.adsabs.harvard.edu/abs/2014ApJ...797...85G}
}

@ARTICLE{2013PASA...30...25D,
       author = {{Dawson}, J.~R.},
        title = "{The Supershell - Molecular Cloud Connection: Large-Scale Stellar Feedback and the Formation of the Molecular ISM}",
      journal = {\pasa},
         year = 2013,
        month = feb,
       volume = {30},
          eid = {e025},
        pages = {e025},
          doi = {10.1017/pas.2013.002},
archivePrefix = {arXiv},
       eprint = {1301.1419},
 primaryClass = {astro-ph.GA},
       adsurl = {https://ui.adsabs.harvard.edu/abs/2013PASA...30...25D}
}

@ARTICLE{1987ApJ...317..190M,
       author = {{McCray}, Richard and {Kafatos}, Minas},
        title = "{Supershells and Propagating Star Formation}",
      journal = {\apj},
         year = 1987,
        month = jun,
       volume = {317},
        pages = {190},
          doi = {10.1086/165267},
       adsurl = {https://ui.adsabs.harvard.edu/abs/1987ApJ...317..190M}
}

@ARTICLE{1994MNRAS.266..567G,
       author = {{Gardiner}, L.~T. and {Sawa}, T. and {Fujimoto}, M.},
        title = "{Numerical simulations of the Magellanic system - I. Orbits of the Magellanic Clouds and the global gas distribution.}",
      journal = {\mnras},
         year = 1994,
        month = feb,
       volume = {266},
        pages = {567-582},
          doi = {10.1093/mnras/266.3.567},
       adsurl = {https://ui.adsabs.harvard.edu/abs/1994MNRAS.266..567G}
}

@ARTICLE{2011ApJ...728..127D,
       author = {{Dawson}, J.~R. and {McClure-Griffiths}, N.~M. and {Kawamura}, A. and {Mizuno}, N. and {Onishi}, T. and {Mizuno}, A. and {Fukui}, Y.},
        title = "{Supershells as Molecular Cloud Factories: Parsec Resolution Observations of H I and $^{12}$CO(J = 1-0) in GSH 287+04-17 and GSH 277+00+36}",
      journal = {\apj},
         year = 2011,
        month = feb,
       volume = {728},
       number = {2},
          eid = {127},
        pages = {127},
          doi = {10.1088/0004-637X/728/2/127},
archivePrefix = {arXiv},
       eprint = {1012.5363},
 primaryClass = {astro-ph.GA},
       adsurl = {https://ui.adsabs.harvard.edu/abs/2011ApJ...728..127D}
}

@ARTICLE{1977ApJ...214..725E,
       author = {{Elmegreen}, B.~G. and {Lada}, C.~J.},
        title = "{Sequential formation of subgroups in OB associations.}",
      journal = {\apj},
         year = 1977,
        month = jun,
       volume = {214},
        pages = {725-741},
          doi = {10.1086/155302},
       adsurl = {https://ui.adsabs.harvard.edu/abs/1977ApJ...214..725E}
}

@ARTICLE{1994MNRAS.268..291W,
       author = {{Whitworth}, A.~P. and {Bhattal}, A.~S. and {Chapman}, S.~J. and {Disney}, M.~J. and {Turner}, J.~A.},
        title = "{The Preferential Formation of High-Mass Stars in Shocked Interstellar Gas Layers}",
      journal = {\mnras},
         year = 1994,
        month = may,
       volume = {268},
        pages = {291},
          doi = {10.1093/mnras/268.1.291},
       adsurl = {https://ui.adsabs.harvard.edu/abs/1994MNRAS.268..291W}
}

@ARTICLE{2022MNRAS.509..272C,
       author = {{Chevance}, M{\'e}lanie and {Kruijssen}, J.~M. Diederik and {Krumholz}, Mark R. and {Groves}, Brent and {Keller}, Benjamin W. and {Hughes}, Annie and {Glover}, Simon C.~O. and {Henshaw}, Jonathan D. and {Herrera}, Cinthya N. and {Kim}, Jaeyeon and {Leroy}, Adam K. and {Pety}, J{\'e}r{\^o}me and {Razza}, Alessandro and {Rosolowsky}, Erik and {Schinnerer}, Eva and {Schruba}, Andreas and {Barnes}, Ashley T. and {Bigiel}, Frank and {Blanc}, Guillermo A. and {Dale}, Daniel A. and {Emsellem}, Eric and {Faesi}, Christopher M. and {Grasha}, Kathryn and {Klessen}, Ralf S. and {Kreckel}, Kathryn and {Liu}, Daizhong and {Longmore}, Steven N. and {Meidt}, Sharon E. and {Querejeta}, Miguel and {Saito}, Toshiki and {Sun}, Jiayi and {Usero}, Antonio},
        title = "{Pre-supernova feedback mechanisms drive the destruction of molecular clouds in nearby star-forming disc galaxies}",
      journal = {\mnras},
         year = 2022,
        month = jan,
       volume = {509},
       number = {1},
        pages = {272-288},
          doi = {10.1093/mnras/stab2938},
archivePrefix = {arXiv},
       eprint = {2010.13788},
 primaryClass = {astro-ph.GA},
       adsurl = {https://ui.adsabs.harvard.edu/abs/2022MNRAS.509..272C}
}

@ARTICLE{2005A&A...433....1A,
       author = {{Audit}, E. and {Hennebelle}, P.},
        title = "{Thermal condensation in a turbulent atomic hydrogen flow}",
      journal = {\aap},
         year = 2005,
        month = apr,
       volume = {433},
       number = {1},
        pages = {1-13},
          doi = {10.1051/0004-6361:20041474},
archivePrefix = {arXiv},
       eprint = {astro-ph/0410062},
 primaryClass = {astro-ph},
       adsurl = {https://ui.adsabs.harvard.edu/abs/2005A&A...433....1A}
}

@ARTICLE{2011ApJ...731...13N,
       author = {{Ntormousi}, Evangelia and {Burkert}, Andreas and {Fierlinger}, Katharina and {Heitsch}, Fabian},
        title = "{Formation of Cold Filamentary Structure from Wind-blown Superbubbles}",
      journal = {\apj},
         year = 2011,
        month = apr,
       volume = {731},
       number = {1},
          eid = {13},
        pages = {13},
          doi = {10.1088/0004-637X/731/1/13},
archivePrefix = {arXiv},
       eprint = {1011.5751},
 primaryClass = {astro-ph.GA},
       adsurl = {https://ui.adsabs.harvard.edu/abs/2011ApJ...731...13N}
}

@ARTICLE{2015MNRAS.453.3190B,
       author = {{Bica}, E. and {Santiago}, B. and {Bonatto}, C. and {Garcia-Dias}, R. and {Kerber}, L. and {Dias}, B. and {Barbuy}, B. and {Balbinot}, E.},
        title = "{Bridge over troubled gas: clusters and associations under the SMC and LMC tidal stresses}",
      journal = {\mnras},
         year = 2015,
        month = nov,
       volume = {453},
       number = {3},
        pages = {3190-3202},
          doi = {10.1093/mnras/stv1720},
archivePrefix = {arXiv},
       eprint = {1507.07725},
 primaryClass = {astro-ph.GA},
       adsurl = {https://ui.adsabs.harvard.edu/abs/2015MNRAS.453.3190B}
}

@ARTICLE{2001MNRAS.326..539G,
       author = {{Graham}, M.~F. and {Smith}, R.~J. and {Meaburn}, J. and {Bryce}, M.},
        title = "{The expansion of the giant filamentary shell, DEM 171, in the Magellanic Bridge}",
      journal = {\mnras},
         year = 2001,
        month = sep,
       volume = {326},
       number = {2},
        pages = {539-542},
          doi = {10.1046/j.1365-8711.2001.04561.x},
       adsurl = {https://ui.adsabs.harvard.edu/abs/2001MNRAS.326..539G}
}

@ARTICLE{1999AJ....118..273W,
       author = {{Walter}, Fabian and {Brinks}, Elias},
        title = "{Holes and Shells in the Interstellar Medium of the Nearby Dwarf Galaxy IC 2574}",
      journal = {\aj},
         year = 1999,
        month = jul,
       volume = {118},
       number = {1},
        pages = {273-301},
          doi = {10.1086/300906},
archivePrefix = {arXiv},
       eprint = {astro-ph/9904002},
 primaryClass = {astro-ph},
       adsurl = {https://ui.adsabs.harvard.edu/abs/1999AJ....118..273W}
}

@ARTICLE{2015ApJ...799...64D,
       author = {{Dawson}, J.~R. and {Ntormousi}, E. and {Fukui}, Y. and {Hayakawa}, T. and {Fierlinger}, K.},
        title = "{A Young Giant Molecular Cloud Formed at the Interface of Two Colliding Supershells: Observations Meet Simulations}",
      journal = {\apj},
         year = 2015,
        month = jan,
       volume = {799},
       number = {1},
          eid = {64},
        pages = {64},
          doi = {10.1088/0004-637X/799/1/64},
archivePrefix = {arXiv},
       eprint = {1411.2708},
 primaryClass = {astro-ph.GA},
       adsurl = {https://ui.adsabs.harvard.edu/abs/2015ApJ...799...64D}
}

@ARTICLE{2022MNRAS.515..940W,
       author = {{Wang}, Jianling and {Hammer}, Francois and {Yang}, Yanbin},
        title = "{Lessons from the Magellanic System and its modeling}",
      journal = {\mnras},
         year = 2022,
        month = sep,
       volume = {515},
       number = {1},
        pages = {940-952},
          doi = {10.1093/mnras/stac1640},
archivePrefix = {arXiv},
       eprint = {2206.04692},
 primaryClass = {astro-ph.GA},
       adsurl = {https://ui.adsabs.harvard.edu/abs/2022MNRAS.515..940W}
}

@ARTICLE{2012MNRAS.421.2109B,
       author = {{Besla}, Gurtina and {Kallivayalil}, Nitya and {Hernquist}, Lars and {van der Marel}, Roeland P. and {Cox}, T.~J. and {Kere{\v{s}}}, Du{\v{s}}an},
        title = "{The role of dwarf galaxy interactions in shaping the Magellanic System and implications for Magellanic Irregulars}",
      journal = {\mnras},
         year = 2012,
        month = apr,
       volume = {421},
       number = {3},
        pages = {2109-2138},
          doi = {10.1111/j.1365-2966.2012.20466.x},
archivePrefix = {arXiv},
       eprint = {1201.1299},
 primaryClass = {astro-ph.GA},
       adsurl = {https://ui.adsabs.harvard.edu/abs/2012MNRAS.421.2109B}
}

@ARTICLE{2023ARA&A..61...19M,
       author = {{McClure-Griffiths}, Naomi M. and {Stanimirovi{\'c}}, Sne{\v{z}}ana and {Rybarczyk}, Daniel R.},
        title = "{Atomic Hydrogen in the Milky Way: A Stepping Stone in the Evolution of Galaxies}",
      journal = {\araa},
         year = 2023,
        month = aug,
       volume = {61},
        pages = {19-63},
          doi = {10.1146/annurev-astro-052920-104851},
archivePrefix = {arXiv},
       eprint = {2307.08464},
 primaryClass = {astro-ph.GA},
       adsurl = {https://ui.adsabs.harvard.edu/abs/2023ARA&A..61...19M}
}

@article{Parker1998,
  author = {Parker, Q. A.},
  title = {AAO Newsletter},
  journal = {AAO Newsletter},
  year = {1998},
  volume = {87},
  pages = {8},
  note = {AAO Newsletter, 87, 8}
}

@ARTICLE{1997MNRAS.289..225S,
       author = {{Staveley-Smith}, L. and {Sault}, R.~J. and {Hatzidimitriou}, D. and {Kesteven}, M.~J. and {McConnell}, D.},
        title = "{An HI aperture synthesis mosaic of the Small Magellanic Cloud}",
      journal = {\mnras},
         year = 1997,
        month = aug,
       volume = {289},
       number = {2},
        pages = {225-252},
          doi = {10.1093/mnras/289.2.225},
       adsurl = {https://ui.adsabs.harvard.edu/abs/1997MNRAS.289..225S}
}

@ARTICLE{2014A&A...564A.116S,
       author = {{Suad}, L.~A. and {Caiafa}, C.~F. and {Arnal}, E.~M. and {Cichowolski}, S.},
        title = "{A new catalog of H i supershell candidates in the outer part of the Galaxy}",
      journal = {\aap},
         year = 2014,
        month = apr,
       volume = {564},
          eid = {A116},
        pages = {A116},
          doi = {10.1051/0004-6361/201323147},
archivePrefix = {arXiv},
       eprint = {1403.4141},
 primaryClass = {astro-ph.GA},
       adsurl = {https://ui.adsabs.harvard.edu/abs/2014A&A...564A.116S}
}

@ARTICLE{2008ApJ...687..303I,
       author = {{Inoue}, Tsuyoshi and {Inutsuka}, Shu-ichiro},
        title = "{Two-Fluid Magnetohydrodynamic Simulations of Converging H I Flows in the Interstellar Medium. I. Methodology and Basic Results}",
      journal = {\apj},
         year = 2008,
        month = nov,
       volume = {687},
       number = {1},
        pages = {303-310},
          doi = {10.1086/590528},
archivePrefix = {arXiv},
       eprint = {0801.0486},
 primaryClass = {astro-ph},
       adsurl = {https://ui.adsabs.harvard.edu/abs/2008ApJ...687..303I}
}

@ARTICLE{2020MNRAS.499.2534K,
       author = {{Kalari}, Venu M. and {Rubio}, Monica and {Salda{\~n}o}, Hugo P. and {Bolatto}, Alberto D.},
        title = "{Resolved star formation in the metal-poor star-forming region Magellanic Bridge C}",
      journal = {\mnras},
         year = 2020,
        month = dec,
       volume = {499},
       number = {2},
        pages = {2534-2553},
          doi = {10.1093/mnras/staa2963},
archivePrefix = {arXiv},
       eprint = {2009.11868},
 primaryClass = {astro-ph.GA},
       adsurl = {https://ui.adsabs.harvard.edu/abs/2020MNRAS.499.2534K}
}

@ARTICLE{5725236,
  author={van der Walt, Stefan and Colbert, S. Chris and Varoquaux, Gael},
  journal={Computing in Science \& Engineering}, 
  title={The NumPy Array: A Structure for Efficient Numerical Computation}, 
  year={2011},
  volume={13},
  number={2},
  pages={22-30},
  doi={10.1109/MCSE.2011.37}}

@ARTICLE{4160265,
  author={Hunter, John D.},
  journal={Computing in Science \& Engineering}, 
  title={Matplotlib: A 2D Graphics Environment}, 
  year={2007},
  volume={9},
  number={3},
  pages={90-95},
  doi={10.1109/MCSE.2007.55}}

@ARTICLE{2020NatMe..17..261V,
       author = {{Virtanen}, Pauli and {Gommers}, Ralf and {Oliphant}, Travis E. and {Haberland}, Matt and {Reddy}, Tyler and {Cournapeau}, David and {Burovski}, Evgeni and {Peterson}, Pearu and {Weckesser}, Warren and {Bright}, Jonathan and {van der Walt}, St{\'e}fan J. and {Brett}, Matthew and {Wilson}, Joshua and {Millman}, K. Jarrod and {Mayorov}, Nikolay and {Nelson}, Andrew R.~J. and {Jones}, Eric and {Kern}, Robert and {Larson}, Eric and {Carey}, C.~J. and {Polat}, {\.I}lhan and {Feng}, Yu and {Moore}, Eric W. and {VanderPlas}, Jake and {Laxalde}, Denis and {Perktold}, Josef and {Cimrman}, Robert and {Henriksen}, Ian and {Quintero}, E.~A. and {Harris}, Charles R. and {Archibald}, Anne M. and {Ribeiro}, Ant{\^o}nio H. and {Pedregosa}, Fabian and {van Mulbregt}, Paul and {SciPy 1. 0 Contributors}},
        title = "{SciPy 1.0: fundamental algorithms for scientific computing in Python}",
      journal = {Nature Methods},
         year = 2020,
        month = feb,
       volume = {17},
        pages = {261-272},
          doi = {10.1038/s41592-019-0686-2},
archivePrefix = {arXiv},
       eprint = {1907.10121},
 primaryClass = {cs.MS},
       adsurl = {https://ui.adsabs.harvard.edu/abs/2020NatMe..17..261V}
}

@ARTICLE{2013A&A...558A..33A,
       author = {{Astropy Collaboration} and {Robitaille}, Thomas P. and {Tollerud}, Erik J. and {Greenfield}, Perry and {Droettboom}, Michael and {Bray}, Erik and {Aldcroft}, Tom and {Davis}, Matt and {Ginsburg}, Adam and {Price-Whelan}, Adrian M. and {Kerzendorf}, Wolfgang E. and {Conley}, Alexander and {Crighton}, Neil and {Barbary}, Kyle and {Muna}, Demitri and {Ferguson}, Henry and {Grollier}, Fr{\'e}d{\'e}ric and {Parikh}, Madhura M. and {Nair}, Prasanth H. and {Unther}, Hans M. and {Deil}, Christoph and {Woillez}, Julien and {Conseil}, Simon and {Kramer}, Roban and {Turner}, James E.~H. and {Singer}, Leo and {Fox}, Ryan and {Weaver}, Benjamin A. and {Zabalza}, Victor and {Edwards}, Zachary I. and {Azalee Bostroem}, K. and {Burke}, D.~J. and {Casey}, Andrew R. and {Crawford}, Steven M. and {Dencheva}, Nadia and {Ely}, Justin and {Jenness}, Tim and {Labrie}, Kathleen and {Lim}, Pey Lian and {Pierfederici}, Francesco and {Pontzen}, Andrew and {Ptak}, Andy and {Refsdal}, Brian and {Servillat}, Mathieu and {Streicher}, Ole},
        title = "{Astropy: A community Python package for astronomy}",
      journal = {\aap},
         year = 2013,
        month = oct,
       volume = {558},
          eid = {A33},
        pages = {A33},
          doi = {10.1051/0004-6361/201322068},
archivePrefix = {arXiv},
       eprint = {1307.6212},
 primaryClass = {astro-ph.IM},
       adsurl = {https://ui.adsabs.harvard.edu/abs/2013A&A...558A..33A}
}

@software{larry_bradley_2024_12585239,
  author       = {Larry Bradley and
                  Brigitta Sip{\H o}cz and
                  Thomas Robitaille and
                  Erik Tollerud and
                  Z\`e Vin{\'{\i}}cius and
                  Christoph Deil and
                  Kyle Barbary and
                  Tom J Wilson and
                  Ivo Busko and
                  Axel Donath and
                  Hans Moritz G{\"u}nther and
                  Mihai Cara and
                  P. L. Lim and
                  Sebastian Me{\ss}linger and
                  Zach Burnett and
                  Simon Conseil and
                  Michael Droettboom and
                  Azalee Bostroem and
                  E. M. Bray and
                  Lars Andersen Bratholm and
                  William Jamieson and
                  Adam Ginsburg and
                  Geert Barentsen and
                  Matt Craig and
                  Sergio Pascual and
                  Shivangee Rathi and
                  Marshall Perrin and
                  Brett M. Morris and
                  Gabriel Perren},
  title        = {astropy/photutils: 1.13.0},
  month        = jun,
  year         = 2024,
  publisher    = {Zenodo},
  version      = {1.13.0},
  doi          = {10.5281/zenodo.12585239},
  url          = {https://doi.org/10.5281/zenodo.12585239
}}

@ARTICLE{2024ARA&A..62..369S,
       author = {{Schinnerer}, E. and {Leroy}, A.~K.},
        title = "{Molecular Gas and the Star-Formation Process on Cloud Scales in Nearby Galaxies}",
      journal = {\araa},
         year = 2024,
        month = sep,
       volume = {62},
       number = {1},
        pages = {369-436},
          doi = {10.1146/annurev-astro-071221-052651},
archivePrefix = {arXiv},
       eprint = {2403.19843},
 primaryClass = {astro-ph.GA},
       adsurl = {https://ui.adsabs.harvard.edu/abs/2024ARA&A..62..369S}
}

@ARTICLE{2011AJ....141...23B,
       author = {{Bagetakos}, I. and {Brinks}, E. and {Walter}, F. and {de Blok}, W.~J.~G. and {Usero}, A. and {Leroy}, A.~K. and {Rich}, J.~W. and {Kennicutt}, Jr., R.~C.},
        title = "{The Fine-scale Structure of the Neutral Interstellar Medium in Nearby Galaxies}",
      journal = {\aj},
         year = 2011,
        month = jan,
       volume = {141},
       number = {1},
          eid = {23},
        pages = {23},
          doi = {10.1088/0004-6256/141/1/23},
archivePrefix = {arXiv},
       eprint = {1008.1845},
 primaryClass = {astro-ph.CO},
       adsurl = {https://ui.adsabs.harvard.edu/abs/2011AJ....141...23B}
}

@ARTICLE{1999A&A...348..728R,
       author = {{Rolleston}, W.~R.~J. and {Dufton}, P.~L. and {McErlean}, N.~D. and {Venn}, K.~A.},
        title = "{The chemical composition of the young, Inter-Cloud population}",
      journal = {\aap},
         year = 1999,
        month = aug,
       volume = {348},
        pages = {728-736},
       adsurl = {https://ui.adsabs.harvard.edu/abs/1999A&A...348..728R}
}

@ARTICLE{2009ApJ...704..161I,
       author = {{Inoue}, Tsuyoshi and {Inutsuka}, Shu-ichiro},
        title = "{Two-Fluid Magnetohydrodynamics Simulations of Converging H I Flows in the Interstellar Medium. II. Are Molecular Clouds Generated Directly from a Warm Neutral Medium?}",
      journal = {\apj},
         year = 2009,
        month = oct,
       volume = {704},
       number = {1},
        pages = {161-169},
          doi = {10.1088/0004-637X/704/1/161},
archivePrefix = {arXiv},
       eprint = {0908.3701},
 primaryClass = {astro-ph.GA},
       adsurl = {https://ui.adsabs.harvard.edu/abs/2009ApJ...704..161I}
}

@ARTICLE{2012ApJ...759...35I,
       author = {{Inoue}, Tsuyoshi and {Inutsuka}, Shu-ichiro},
        title = "{Formation of Turbulent and Magnetized Molecular Clouds via Accretion Flows of H I Clouds}",
      journal = {\apj},
         year = 2012,
        month = nov,
       volume = {759},
       number = {1},
          eid = {35},
        pages = {35},
          doi = {10.1088/0004-637X/759/1/35},
archivePrefix = {arXiv},
       eprint = {1205.6217},
 primaryClass = {astro-ph.GA},
       adsurl = {https://ui.adsabs.harvard.edu/abs/2012ApJ...759...35I}
}

@ARTICLE{2015A&A...580A..49I,
       author = {{Inutsuka}, Shu-ichiro and {Inoue}, Tsuyoshi and {Iwasaki}, Kazunari and {Hosokawa}, Takashi},
        title = "{The formation and destruction of molecular clouds and galactic star formation. An origin for the cloud mass function and star formation efficiency}",
      journal = {\aap},
         year = 2015,
        month = aug,
       volume = {580},
          eid = {A49},
        pages = {A49},
          doi = {10.1051/0004-6361/201425584},
archivePrefix = {arXiv},
       eprint = {1505.04696},
 primaryClass = {astro-ph.GA},
       adsurl = {https://ui.adsabs.harvard.edu/abs/2015A&A...580A..49I}
}

@ARTICLE{2008ApJ...678..219L,
       author = {{Lehner}, N. and {Howk}, J.~C. and {Keenan}, F.~P. and {Smoker}, J.~V.},
        title = "{Metallicity and Physical Conditions in the Magellanic Bridge}",
      journal = {\apj},
         year = 2008,
        month = may,
       volume = {678},
       number = {1},
        pages = {219-233},
          doi = {10.1086/529574},
archivePrefix = {arXiv},
       eprint = {0801.2534},
 primaryClass = {astro-ph},
       adsurl = {https://ui.adsabs.harvard.edu/abs/2008ApJ...678..219L}
}

@ARTICLE{2012MNRAS.421.3159M,
       author = {{Moss}, V.~A. and {McClure-Griffiths}, N.~M. and {Braun}, R. and {Hill}, A.~S. and {Madsen}, G.~J.},
        title = "{GSH 006-15+7: a local Galactic supershell featuring transition from H I emission to absorption}",
      journal = {\mnras},
         year = 2012,
        month = apr,
       volume = {421},
       number = {4},
        pages = {3159-3169},
          doi = {10.1111/j.1365-2966.2012.20538.x},
archivePrefix = {arXiv},
       eprint = {1201.2700},
 primaryClass = {astro-ph.GA},
       adsurl = {https://ui.adsabs.harvard.edu/abs/2012MNRAS.421.3159M}
}

@ARTICLE{2010A&A...521A..17K,
       author = {{Kalberla}, P.~M.~W. and {McClure-Griffiths}, N.~M. and {Pisano}, D.~J. and {Calabretta}, M.~R. and {Ford}, H. Alyson and {Lockman}, F.~J. and {Staveley-Smith}, L. and {Kerp}, J. and {Winkel}, B. and {Murphy}, T. and {Newton-McGee}, K.},
        title = "{GASS: the Parkes Galactic all-sky survey. II. Stray-radiation correction and second data release}",
      journal = {\aap},
         year = 2010,
        month = oct,
       volume = {521},
          eid = {A17},
        pages = {A17},
          doi = {10.1051/0004-6361/200913979},
archivePrefix = {arXiv},
       eprint = {1007.0686},
 primaryClass = {astro-ph.GA},
       adsurl = {https://ui.adsabs.harvard.edu/abs/2010A&A...521A..17K}
}

@ARTICLE{2007MNRAS.381L..11M,
       author = {{Muller}, Erik and {Bekki}, Kenji},
        title = "{The origin of large-scale HI structures in the Magellanic Bridge}",
      journal = {\mnras},
         year = 2007,
        month = oct,
       volume = {381},
       number = {1},
        pages = {L11-L15},
          doi = {10.1111/j.1745-3933.2007.00356.x},
archivePrefix = {arXiv},
       eprint = {0706.3544},
 primaryClass = {astro-ph},
       adsurl = {https://ui.adsabs.harvard.edu/abs/2007MNRAS.381L..11M}
}

@ARTICLE{2014AJ....147..122D,
       author = {{de Grijs}, Richard and {Wicker}, James E. and {Bono}, Giuseppe},
        title = "{Clustering of Local Group Distances: Publication Bias or Correlated Measurements? I. The Large Magellanic Cloud}",
      journal = {\aj},
         year = 2014,
        month = may,
       volume = {147},
       number = {5},
          eid = {122},
        pages = {122},
          doi = {10.1088/0004-6256/147/5/122},
archivePrefix = {arXiv},
       eprint = {1403.3141},
 primaryClass = {astro-ph.GA},
       adsurl = {https://ui.adsabs.harvard.edu/abs/2014AJ....147..122D}
}

@ARTICLE{2015AJ....149..179D,
       author = {{de Grijs}, Richard and {Bono}, Giuseppe},
        title = "{Clustering of Local Group Distances: Publication Bias or Correlated Measurements? III. The Small Magellanic Cloud}",
      journal = {\aj},
         year = 2015,
        month = jun,
       volume = {149},
       number = {6},
          eid = {179},
        pages = {179},
          doi = {10.1088/0004-6256/149/6/179},
archivePrefix = {arXiv},
       eprint = {1504.00417},
 primaryClass = {astro-ph.SR},
       adsurl = {https://ui.adsabs.harvard.edu/abs/2015AJ....149..179D}
}

@ARTICLE{2017ApJ...839..107P,
       author = {{Pineda}, Jorge L. and {Langer}, William D. and {Goldsmith}, Paul F. and {Horiuchi}, Shinji and {Kuiper}, Thomas B.~H. and {Muller}, Erik and {Hughes}, Annie and {Ott}, J{\"u}rgen and {Requena-Torres}, Miguel A. and {Velusamy}, Thangasamy and {Wong}, Tony},
        title = "{Characterizing the Transition from Diffuse Atomic to Dense Molecular Clouds in the Magellanic Clouds with [C II], [C I], and CO}",
      journal = {\apj},
         year = 2017,
        month = apr,
       volume = {839},
       number = {2},
          eid = {107},
        pages = {107},
          doi = {10.3847/1538-4357/aa683a},
archivePrefix = {arXiv},
       eprint = {1704.00739},
 primaryClass = {astro-ph.GA},
       adsurl = {https://ui.adsabs.harvard.edu/abs/2017ApJ...839..107P}
}

@ARTICLE{2022ARA&A..60..319S,
       author = {{Saintonge}, Am{\'e}lie and {Catinella}, Barbara},
        title = "{The Cold Interstellar Medium of Galaxies in the Local Universe}",
      journal = {\araa},
         year = 2022,
        month = aug,
       volume = {60},
        pages = {319-361},
          doi = {10.1146/annurev-astro-021022-043545},
archivePrefix = {arXiv},
       eprint = {2202.00690},
 primaryClass = {astro-ph.GA},
       adsurl = {https://ui.adsabs.harvard.edu/abs/2022ARA&A..60..319S}
}

@ARTICLE{2011AJ....142...37S,
       author = {{Schruba}, Andreas and {Leroy}, Adam K. and {Walter}, Fabian and {Bigiel}, Frank and {Brinks}, Elias and {de Blok}, W.~J.~G. and {Dumas}, Gaelle and {Kramer}, Carsten and {Rosolowsky}, Erik and {Sandstrom}, Karin and {Schuster}, Karl and {Usero}, Antonio and {Weiss}, Axel and {Wiesemeyer}, Helmut},
        title = "{A Molecular Star Formation Law in the Atomic-gas-dominated Regime in Nearby Galaxies}",
      journal = {\aj},
         year = 2011,
        month = aug,
       volume = {142},
       number = {2},
          eid = {37},
        pages = {37},
          doi = {10.1088/0004-6256/142/2/37},
archivePrefix = {arXiv},
       eprint = {1105.4605},
 primaryClass = {astro-ph.CO},
       adsurl = {https://ui.adsabs.harvard.edu/abs/2011AJ....142...37S}
}

@ARTICLE{2003ApJ...587..278W,
       author = {{Wolfire}, Mark G. and {McKee}, Christopher F. and {Hollenbach}, David and {Tielens}, A.~G.~G.~M.},
        title = "{Neutral Atomic Phases of the Interstellar Medium in the Galaxy}",
      journal = {\apj},
         year = 2003,
        month = apr,
       volume = {587},
       number = {1},
        pages = {278-311},
          doi = {10.1086/368016},
archivePrefix = {arXiv},
       eprint = {astro-ph/0207098},
 primaryClass = {astro-ph},
       adsurl = {https://ui.adsabs.harvard.edu/abs/2003ApJ...587..278W}
}

@ARTICLE{1995ApJ...443..152W,
       author = {{Wolfire}, M.~G. and {Hollenbach}, D. and {McKee}, C.~F. and {Tielens}, A.~G.~G.~M. and {Bakes}, E.~L.~O.},
        title = "{The Neutral Atomic Phases of the Interstellar Medium}",
      journal = {\apj},
         year = 1995,
        month = apr,
       volume = {443},
        pages = {152},
          doi = {10.1086/175510},
       adsurl = {https://ui.adsabs.harvard.edu/abs/1995ApJ...443..152W}
}

@ARTICLE{2020A&A...643A.141M,
       author = {{Madden}, S.~C. and {Cormier}, D. and {Hony}, S. and {Lebouteiller}, V. and {Abel}, N. and {Galametz}, M. and {De Looze}, I. and {Chevance}, M. and {Polles}, F.~L. and {Lee}, M.-Y. and {Galliano}, F. and {Lambert-Huyghe}, A. and {Hu}, D. and {Ramambason}, L.},
        title = "{Tracing the total molecular gas in galaxies: [CII] and the CO-dark gas}",
      journal = {\aap},
         year = 2020,
        month = nov,
       volume = {643},
          eid = {A141},
        pages = {A141},
          doi = {10.1051/0004-6361/202038860},
archivePrefix = {arXiv},
       eprint = {2009.00649},
 primaryClass = {astro-ph.GA},
       adsurl = {https://ui.adsabs.harvard.edu/abs/2020A&A...643A.141M}
}

@ARTICLE{1999MNRAS.302..417S,
       author = {{Stanimirovic}, S. and {Staveley-Smith}, L. and {Dickey}, J.~M. and {Sault}, R.~J. and {Snowden}, S.~L.},
        title = "{The large-scale HI structure of the Small Magellanic Cloud}",
      journal = {\mnras},
         year = 1999,
        month = jan,
       volume = {302},
       number = {3},
        pages = {417-436},
          doi = {10.1046/j.1365-8711.1999.02013.x},
       adsurl = {https://ui.adsabs.harvard.edu/abs/1999MNRAS.302..417S}
}

@ARTICLE{2012MNRAS.422.2291P,
       author = {{Parkin}, T.~J. and {Wilson}, C.~D. and {Foyle}, K. and {Baes}, M. and {Bendo}, G.~J. and {Boselli}, A. and {Boquien}, M. and {Cooray}, A. and {Cormier}, D. and {Davies}, J.~I. and {Eales}, S.~A. and {Galametz}, M. and {Gomez}, H.~L. and {Lebouteiller}, V. and {Madden}, S. and {Mentuch}, E. and {Page}, M.~J. and {Pohlen}, M. and {Remy}, A. and {Roussel}, H. and {Sauvage}, M. and {Smith}, M.~W.~L. and {Spinoglio}, L.},
        title = "{The gas-to-dust mass ratio of Centaurus A as seen by Herschel}",
      journal = {\mnras},
         year = 2012,
        month = may,
       volume = {422},
       number = {3},
        pages = {2291-2301},
          doi = {10.1111/j.1365-2966.2012.20778.x},
archivePrefix = {arXiv},
       eprint = {1202.5323},
 primaryClass = {astro-ph.CO},
       adsurl = {https://ui.adsabs.harvard.edu/abs/2012MNRAS.422.2291P}
}

@ARTICLE{2015ApJ...799...96G,
       author = {{Groves}, Brent A. and {Schinnerer}, Eva and {Leroy}, Adam and {Galametz}, Maud and {Walter}, Fabian and {Bolatto}, Alberto and {Hunt}, Leslie and {Dale}, Daniel and {Calzetti}, Daniela and {Croxall}, Kevin and {Kennicutt}, Jr., Robert},
        title = "{Dust Continuum Emission as a Tracer of Gas Mass in Galaxies}",
      journal = {\apj},
         year = 2015,
        month = jan,
       volume = {799},
       number = {1},
          eid = {96},
        pages = {96},
          doi = {10.1088/0004-637X/799/1/96},
archivePrefix = {arXiv},
       eprint = {1411.2975},
 primaryClass = {astro-ph.GA},
       adsurl = {https://ui.adsabs.harvard.edu/abs/2015ApJ...799...96G}
}

@ARTICLE{1977ApJ...218..377W,
       author = {{Weaver}, R. and {McCray}, R. and {Castor}, J. and {Shapiro}, P. and {Moore}, R.},
        title = "{Interstellar bubbles. II. Structure and evolution.}",
      journal = {\apj},
         year = 1977,
        month = dec,
       volume = {218},
        pages = {377-395},
          doi = {10.1086/155692},
       adsurl = {https://ui.adsabs.harvard.edu/abs/1977ApJ...218..377W}
}

@ARTICLE{2026MNRAS.549ag863D,
       author = {{Dempsey}, James and {McClure-Griffiths}, N.~M. and {Marchal}, Antoine and {Clark}, S.~E. and {Dickey}, John M. and {Lee}, Min-Young and {Murray}, Claire and {Nguyen}, Hiep and {Pingel}, Nickolas M. and {Stanimirovi{\'c}}, Sne{\v{z}}ana and {van Loon}, Jacco Th and {D{\'e}nes}, Helga and {Gibson}, Steven J. and {Jameson}, Katie and {Kemp}, Ian and {Lynn}, Callum and {Ma}, Yik Ki},
        title = "{Revealing the cold skeleton of the Magellanic clouds and the Magellanic Bridge with ASKAP}",
      journal = {\mnras},
         year = 2026,
        month = jun,
       volume = {549},
       number = {1},
          eid = {stag863},
        pages = {stag863},
          doi = {10.1093/mnras/stag863},
archivePrefix = {arXiv},
       eprint = {2605.04632},
 primaryClass = {astro-ph.GA},
       adsurl = {https://ui.adsabs.harvard.edu/abs/2026MNRAS.549ag863D}
}

@ARTICLE{1981MNRAS.196..101B,
       author = {{Barlow}, M.~J. and {Smith}, L.~J. and {Willis}, A.~J.},
        title = "{Mass-loss rates for 21 Wolf-rayet stars.}",
      journal = {\mnras},
         year = 1981,
        month = jul,
       volume = {196},
        pages = {101-110},
          doi = {10.1093/mnras/196.2.101},
       adsurl = {https://ui.adsabs.harvard.edu/abs/1981MNRAS.196..101B}
}

@ARTICLE{2012ApJ...745..173W,
       author = {{Welty}, Daniel E. and {Xue}, Rui and {Wong}, Tony},
        title = "{Interstellar H I and H$_{2}$ in the Magellanic Clouds: An Expanded Sample Based on Ultraviolet Absorption-line Data}",
      journal = {\apj},
         year = 2012,
        month = feb,
       volume = {745},
       number = {2},
          eid = {173},
        pages = {173},
          doi = {10.1088/0004-637X/745/2/173},
archivePrefix = {arXiv},
       eprint = {1111.3674},
 primaryClass = {astro-ph.GA},
       adsurl = {https://ui.adsabs.harvard.edu/abs/2012ApJ...745..173W}
}

@ARTICLE{2009ApJ...693..216K,
       author = {{Krumholz}, Mark R. and {McKee}, Christopher F. and {Tumlinson}, Jason},
        title = "{The Atomic-to-Molecular Transition in Galaxies. II: H I and H$_{2}$ Column Densities}",
      journal = {\apj},
         year = 2009,
        month = mar,
       volume = {693},
       number = {1},
        pages = {216-235},
          doi = {10.1088/0004-637X/693/1/216},
archivePrefix = {arXiv},
       eprint = {0811.0004},
 primaryClass = {astro-ph},
       adsurl = {https://ui.adsabs.harvard.edu/abs/2009ApJ...693..216K}
}



\appendix
\section{Gaussian components of the subtracted foreground emission}
\label{app:foreground}
\begin{figure}
    \includegraphics[width=\columnwidth]{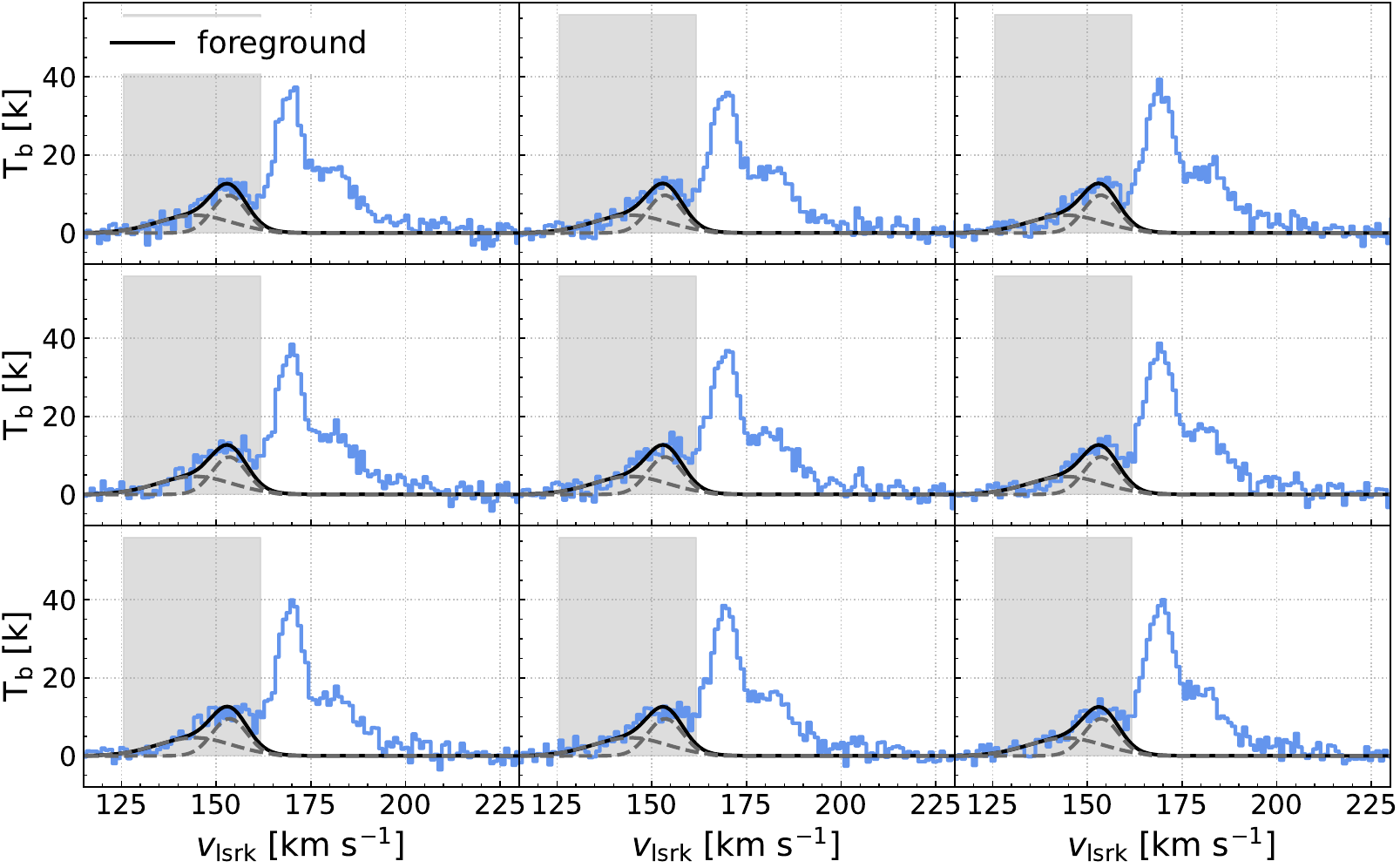}
    \caption{Example of nine neighboring spectra selected to highlight the foreground/shell emission observed in the data. 
    The solid blue lines show the observed data. 
    The gray dashed lines represent the Gaussian models fitted to foreground components, and the solid black line the summed model. 
    The gray shaded area indicates the velocity range where the foreground emission dominates.
    }
    \label{fig:HI_shell_mosaic_spectrum_foreground}
\end{figure}

\begin{figure}
    \centering
    \includegraphics[width=0.9\linewidth]{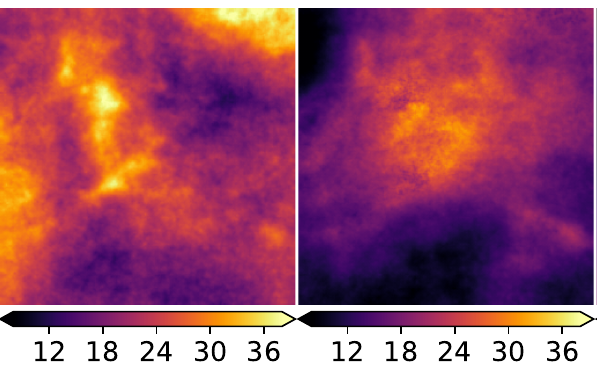}
    \caption{Column density maps (in units of 10$^{19}$\,cm$^{-2}$) of the two Gaussian components in low velocity channels ($v < $ 160\,\kms) of H{\sc i} Shell-91 likely unrelated to shell structures.}
    \label{fig:filament}
\end{figure}

In support of Section~\ref{sec:foreground_sub}, we provide in Figure~\ref{fig:HI_shell_mosaic_spectrum_foreground} an example of nine neighboring spectra showing the two fitted components used to subtract the foreground emission from the data cube centered on Shell-91. The gray shaded area indicates the velocity range where the foreground emission dominates. 
In addition, we also provide in Figure~\ref{fig:filament}, the column density maps of these two components that show no evidence of shell-like morphology.

\section{Cold H{\sc i} column density map and its uncertainties}\label{app:uncertainties}

We assess the impact of {\sc ROHSA} parameter values on the spatial distribution and small-scale clustering of the cold H{\sc i} column density, as shown in Figure~\ref{fig:three_plots}. Following the methodology described in \citet{2022ApJ...937...81T}, we estimate the {\tt ROHSA} uncertainties by performing three series of {\tt ROHSA} Gaussian decompositions, each consisting of 50 runs. 

The first series explores the effect of observational noise by injecting 50 different 3D random noises into the {\tt ROHSA} Gaussian model. For the second series, we perturb the four hyper-parameter values by $\pm$ 10$\%$ of their original values. The third series explores the effect of initial Gaussian guesses by initializing {\tt ROHSA} with 50 randomly selected pixels from the original $\sigma-\mu$ diagrams. For each series, we keep the other parameters fixed. 

For each run, we examine the $\sigma-\mu$ space, confirming the stability of the ``V" shape in $\sigma-\mu$ space shown in Figure~\ref{fig:sigma_mu_diagram}, with the G$_2$ component consistently exhibiting the narrowest velocity dispersions. We then classify the cold H{\sc i} gas components based on velocity dispersions (with $\sigma$ < 2.5\,\kms) within the G$_2$ component. For each series, we obtain the mean column density maps and standard deviation maps. Among the three series, the third shows a slightly higher standard deviations than the other two. 

\begin{figure}
    \centering
    \includegraphics[width=\linewidth]{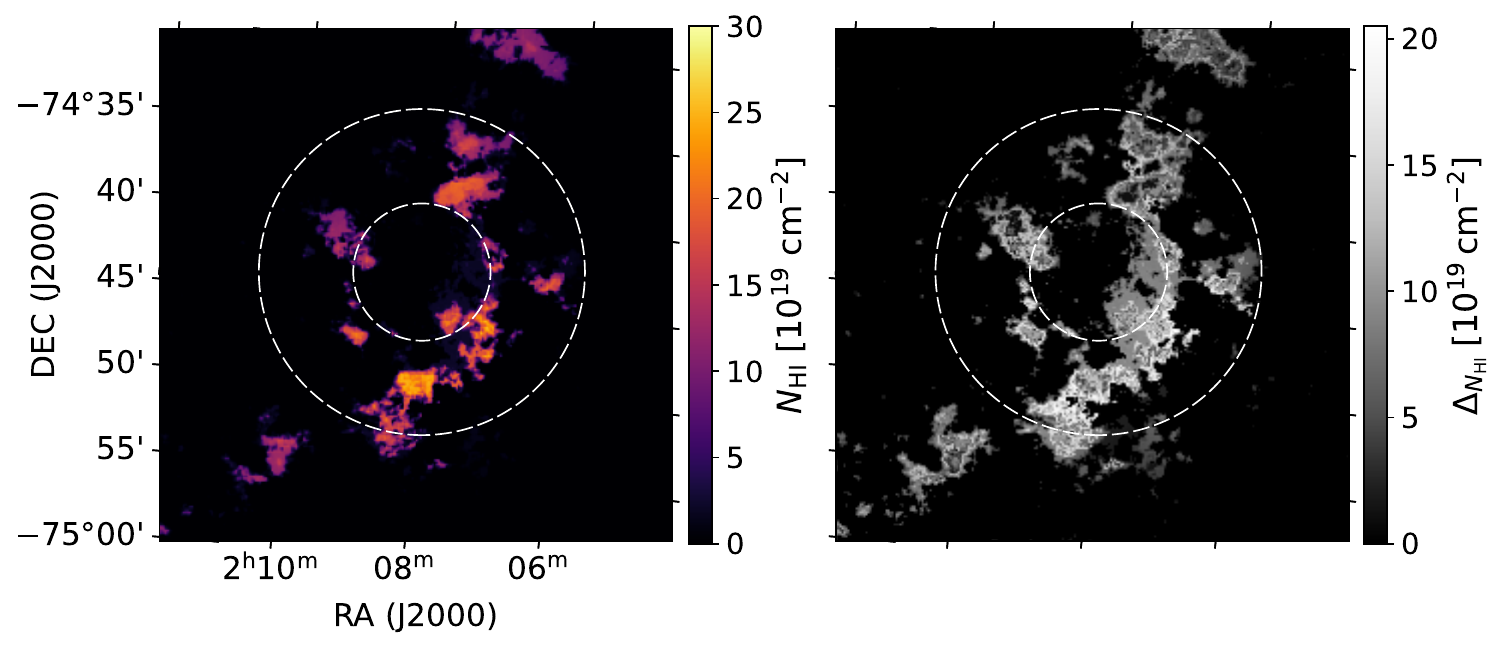}
    \caption{(Left) Mean column density map of cold H{\sc i} gas with $\sigma < 2.5\,\kms$, averaged over {\tt ROHSA} runs with different parameters. (Right) Corresponding uncertainty map. Two white dashed circles denote the inner and outer boundary of Shell-91, obtained from the H{\sc i} channel maps described in Section~\ref{sec:3.1}.
    }
    \label{fig:NHI_cold_with_std}
\end{figure}

Finally, we obtain the cold gas column density and uncertainty maps by averaging the mean column density maps from the three series and combining the standard deviations from each series in quadrature. We display the final results in Figure~\ref{fig:NHI_cold_with_std}: the left panel shows the mean column density map of cold components of H{\sc i} Shell-91, while the right panel shows its corresponding uncertainty map. While the uncertainty map shows that the cold $N_{\rm HI}$ values may differ across each decompositon, the overall cold H{\sc i} column density distributions, found within shell and loop structures, are robust against variations in {\tt ROHSA} parameters.

\bsp	
\label{lastpage}
\end{document}